\documentstyle[preprint,fancybox,epsf,amssymb,amsbsy,amstext,subeqnarray,amsfonts,aps]{revtex}
\draft
\tighten
\begin{document}

\def\kdagger{k\hspace{-0.3cm}/}
\def\kpdagger{k^{'}\hspace{-0.37cm}/ \hspace{.1cm}}
\def\ppidagger{p_\pi \hspace{-0.39cm}/ \hspace{.11cm}}
\def\pdeldagger{p_\Delta \hspace{-0.45cm}/ \hspace{.18cm}}
\def\pindagger{p_N\hspace{-0.46cm}/ \hspace{.18cm}}
\def\poutdagger{p_N^{\, '}\hspace{-0.48cm}/ \hspace{.2cm}}

\title{Magnetic dipole moment of the $\Delta^+$(1232) \\
from the $\gamma \, p \, \to \, \gamma \, \pi^0 \, p$ reaction}
\normalsize
\author{ {D. Drechsel and M. Vanderhaeghen} }
\address{Institut f\"ur Kernphysik, Johannes-Gutenberg-Universit\"at, D-55099 Mainz, Germany}
\date{\today}
\maketitle

\begin{abstract}
The $\gamma p \to \gamma \pi^0 p$ reaction in the
$\Delta(1232)$-resonance 
region is investigated as a method to access the
$\Delta^+(1232)$ magnetic dipole moment. The calculations are
performed within the context of an effective Lagrangian model 
containing both the $\Delta$-resonant mechanism 
and a background of non-resonant contributions to the 
$\gamma p \to \gamma \pi^0 p$ reaction. Results are shown both for 
existing and forthcoming $\gamma p \to \gamma \pi^0 p$ experiments. 
In particular, the sensitivity of unpolarized cross sections and
photon asymmetries to the $\Delta^+$ magnetic dipole moment is displayed for
those forthcoming data. 
\end{abstract}

\section{Introduction}
\label{introduction}

The static properties of baryons constitute an important test for
theoretical descriptions in the non-perturbative domain of QCD. In
particular, different predictions for the magnetic moments 
of baryons were made in quark-models, 
chiral quark soliton models, and lattice QCD calculations 
(see e.g. Ref.~\cite{Ali00}).
\newline
\indent
The experimental values for the magnetic moments of octet baryons ($N,
\Lambda, \Sigma, \Xi$) are known accurately through spin precession
measurements. For decuplet baryons, only the magnetic moment of the
$\Omega^-$ could be determined through such techniques, because the
other decuplet baryons have too short lifetimes. 
\newline
\indent
In particular the magnetic dipole moment of the $\Delta(1232)$ resonance is
of considerable theoretical interest. If SU(6) symmetry holds, then
the nucleon and the $\Delta$ resonance are degenerate 
and their magnetic moments are
related through $\mu_\Delta = e_\Delta \, \mu_p$, where $e_\Delta$ is
the $\Delta$ electric charge, and $\mu_p$ the proton magnetic
moment. However, different theoretical models predict considerable 
deviations from this SU(6) value \cite{Ali00b}.
\newline
\indent
In the past, it has been proposed to determine the magnetic moment of
the $\Delta^{++}(1232)$ by measuring the 
$\pi^+ p \to \gamma \pi^+ p$ reaction \cite{Kon68}, and 
the first dedicated experiment of this type  
has been performed in the '70s \cite{Nef78}.
As a result of these measurements \cite{Nef78,Bos91}, 
and using different theoretical
analyses, the PDG \cite{PDG00} quotes the range~:
$\mu_{\Delta^{++}} = 3.7 - 7.5 \,
\mu_N$ (where $\mu_N$ is the nuclear magneton), while  
SU(6) symmetry results in the value $\mu_{\Delta^{++}} = 5.58 \, \mu_N$. 
The large uncertainty in the extraction of $\mu_{\Delta^{++}}$ from the data 
is due to large non-resonant contributions to the 
$\pi^+ p \to \gamma \pi^+ p$ reaction because of bremsstrahlung from the
charged pion ($\pi^+$) and proton (p). 
\newline
\indent
As an alternative method, it has been proposed \cite{Dre83} 
to determine the magnetic moment of the $\Delta^+(1232)$ 
through measurement of the $\gamma p \to \gamma \pi^0 p$ reaction.  
First calculations for this reaction, including only the resonant
mechanism, have been recently performed in Refs.~\cite{Mac99,DVGS}. 
Due to the small cross sections for this reaction, which is
proportional to $\alpha_{em}^2 = 1/(137)^2$, 
a first measurement has only now been performed by the A2/TAPS
collaboration at MAMI \cite{Kot99}. 
In the near future, such experiments can be performed with much higher
count rates by using $4 \pi$ detectors, such as planned e.g. by using
the Crystal Ball detector at MAMI \cite{CB}.
\newline
\indent
It is the aim of the present work to perform detailed
calculations of the 
 $\gamma \, p \, \to \, \gamma^{\, '} \, \pi^0 \, p$ reaction, which
 include both resonant and non-resonant mechanisms, and to
 investigate in detail the sensitivity of this reaction to the 
$\Delta^+(1232)$ magnetic dipole moment.
\newline
\indent
In Sec.~\ref{kinem}, we start by specifying the kinematics and cross
section of the $\gamma p \to \gamma \pi^0 p$ reaction. 
\newline
\indent
In Sec.~\ref{model}, we describe in detail the effective Lagrangian
model for the $\gamma p \to \gamma \pi^0 p$ reaction in the
$\Delta(1232)$ resonance region. Besides resonant mechanisms which
contain the $\Delta^+$ magnetic moment contribution, 
the model also includes the main non-resonant contributions to this reaction. 
It is shown how the model for the $\gamma p \to \gamma
\pi^0 p$ process is obtained from an analogous model for the $\gamma
p \to \pi^0 p$ process by coupling a photon in a gauge invariant way to
this latter process. Special attention is paid to the constraint which
gauge invariance imposes on the electromagnetic coupling of a photon to
a resonance of finite width, which has been discussed extensively 
in the literature in recent years for the analogous case of the 
$W^\pm$ gauge boson for which very accurate data have become available. 
\newline
\indent
We then show in Sec.~\ref{results}, within the context of the
developed effective Lagrangian model, the results for the $\gamma p
\to \gamma \pi^0 p$ reaction, for both existing and forthcoming
experiments. In particular, we study the sensitivity of 
both unpolarized cross sections and photon asymmetries 
to the $\Delta^+$ magnetic dipole moment. 
\newline
\indent
Finally, we present our conclusions and an outlook in Sec.~\ref{conclusions}.

\section{Kinematics and cross section for 
the $\gamma \, p \to  \gamma \, \pi^0 \, p$ reaction}
\label{kinem}

In the $\gamma p \to \gamma \pi^0 p$ process, 
a photon (with four-momentum $k$ and helicity $\lambda$) 
scatters on a proton target (with four-momentum $p_N$, and spin projection
$s_N$), and a photon (with four-momentum $k'$ and helicity
$\lambda'$), a $\pi^0$ (with four-momentum $p_\pi$) 
and a proton (with four-momentum $p_N^{'}$, spin projection $s'_N$)
can be observed in the final state.
\newline
\indent
The kinematics of the $\gamma p \to \gamma \pi^0 p$ reaction is
characterized by 5 variables. First, the total c.m. energy squared
is given by the usual Mandelstam invariant $s = (k + p_N)^2$. 
Equivalently to $s$, we will use the energy of the intial photon 
$ E_\gamma^{c.m.}$ in the c.m. system of the intial 
$\gamma p$ system (total c.m. system).
Furthermore, there are two variables which characterize the photon 
in the total c.m. system~: its energy $ E_\gamma^{\, ' c.m.}$ and polar angle
$\Theta_\gamma^{c.m.}$ respectively. 
Finally, two more variables are needed to specify the kinematics of
the 3-body final state completely. For convenience, we choose them to
be the pion angles in the rest frame of the 
$\pi^0 p$ system w.r.t. the direction
of the c.m. momentum vector of the $\pi^0 p$ system : $\Theta_\pi^{*}$
denotes the polar angle of the pion, and $\Phi_\pi^{*}$ is its azimuthal
angle. 
From these variables, one can reconstruct the invariant 
mass of the $\pi^0 p$ system as~:
\begin{equation}
W_{\pi^0 p}^2 \,=\, s \,-\,2 \, \sqrt{s} \, E_\gamma^{\, ' c.m.} \, ,
\end{equation}
and the pion momentum in the  $\pi^0 p$  rest frame is calculated as~:
\begin{equation}
| \vec p_\pi^{\, *} |^2 \,=\, {1 \over {4 \,W_{\pi^0 p}^2 }} \,
\left\{ W_{\pi^0 p}^4 \,-\, 2 \, W_{\pi^0 p}^2 \, 
\Big( m_\pi^2 + M_N^2 \Big) \,+\, \Big( M_N^2 - m_\pi^2 \Big)^2
\right\} \, .
\end{equation}
\newline
\indent
The unpolarized cross section for the $\gamma p \to \gamma \pi^0 p$ reaction,
which is differential w.r.t. the outgoing photon energy and
angles in the c.m. system, and differential w.r.t. the pion angles in
the $\pi^0 p$ rest frame is given by~: 
\begin{eqnarray}
{{d \sigma} \over {d E_\gamma^{\, ' c.m.} \, d \Omega_\gamma^{c.m.} \,
    \ d \Omega_\pi^*}} \;&=&\; {1 \over {(2 \pi)^5}} \, {1 \over {32
    \sqrt{s}}} \, {{E_\gamma^{\, ' c.m.}} \over {E_\gamma^{c.m.}}} \,
{{| \vec p_\pi^{\, *} | }\over {W_{\pi^0 p}}} \, \nonumber\\ 
&\times& 
\left(\, {1 \over 4} \sum_{\lambda} \sum_{s_N} \sum_{\lambda'} \sum_{s_N'} 
| \, \varepsilon_\mu(k, \lambda) \, \varepsilon_\nu^*(k', \lambda') 
\, {\mathcal M}^{\nu \mu} \, |^2 \, \right) \, ,
\label{eq:cross3b}
\end{eqnarray}
where $\varepsilon_\mu(k, \lambda)$ ($\varepsilon_\nu^*(k', \lambda')$) are 
the polarization vectors of the incoming (outgoing) photons respectively.  
In Eq.~(\ref{eq:cross3b}), ${\mathcal M}^{\nu \mu}$ is the 
tensor for the $\gamma p \to \gamma \pi^0 p$ process which will
be calculated in Sec.~\ref{model} within the context of an effective
Lagrangian model. 
\newline
\indent
Further on, we will show results for partially integrated cross
sections of the $\gamma p \to \gamma \pi^0 p$ reaction, 
e.g. the cross section $d \sigma / d E_\gamma^{\, ' c.m.}$
differential w.r.t. the outgoing photon c.m. energy, or 
the cross section $d \sigma / d \Omega_\gamma^{c.m.}$ differential
w.r.t. the photon c.m. solid angle. 
Those are obtained by integrating the fully differential cross section
of Eq.~(\ref{eq:cross3b}) over the appropriate variables.

\section{Effective Lagrangian model for the 
$\gamma \, p \to  \gamma \, \pi^0 \, p$ reaction 
in the $\Delta$(1232) resonance region}
\label{model}

\subsection{$\Delta(1232)$ resonance processes}

\begin{figure}[ht]
\epsfxsize=8 cm
\centerline{\epsffile{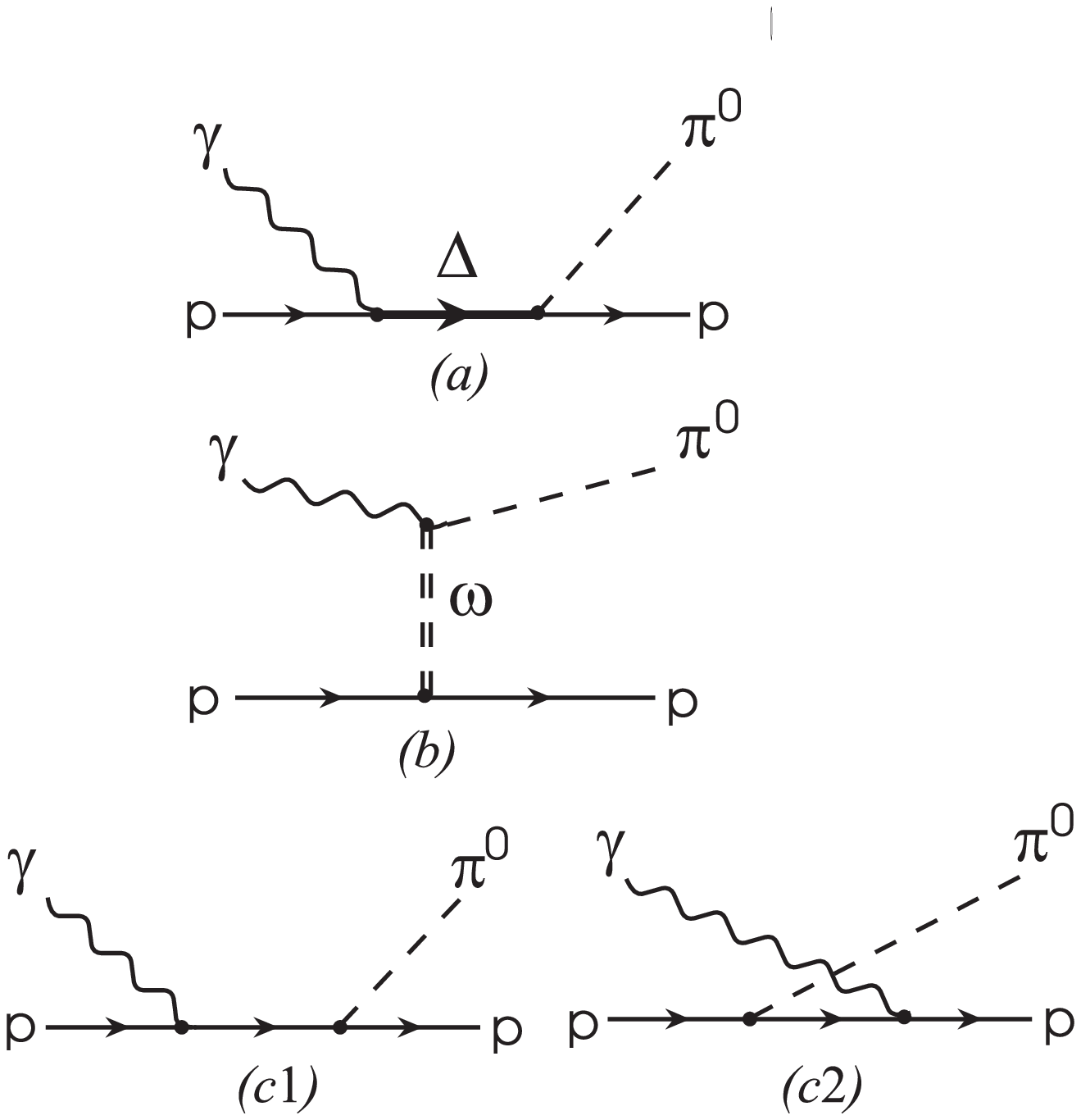}}
\vspace{.5cm}
\caption[]{Diagrams for the $\gamma \, p \, \to \, \pi^0 \, p$ 
reaction in the $\Delta(1232)$ region, 
serving as starting point for the calculation of the 
$\gamma \, p \, \to \, \gamma \, \pi^0 \, p$ reaction~: 
$\Delta$-resonance excitation (a),
$\omega$-exchange (b), and Born diagrams (c1,c2).}
\label{fig:gap_piop_diag}
\end{figure}

In order to calculate the resonant $\Delta(1232)$ contribution 
to the $\gamma p \to \gamma \pi^0 p$ reaction, we start from the
$\Delta(1232)$-resonance contribution to the $\gamma p \to \pi^0 p$
process, corresponding to Fig.~\ref{fig:gap_piop_diag}(a).  
\newline
\indent
We describe the $\Delta$-resonance in the Rarita-Schwinger formalism. 
It is well known that the description of a spin-3/2 field, such as the
$\Delta$-resonance, through a Rarita-Schwinger field, introduces an
arbitrary parameter (usually called $A$ in the literature, see e.g. 
\cite{Nath71}). This parameter arises when removing the spin-1/2
component of the vector-spinor Rarita-Schwinger field, through a
contact transformation. Physical quantities, describing a spin-3/2
particle, should however be independent of this parameter $A$ and 
therefore care has to be taken in the calculations that the $A$-dependence
of the $\Delta$ vertices ($\gamma N \Delta$, $\pi N \Delta$) and
$\Delta$ propagator cancel. We follow here the procedure of
Refs.~\cite{Ami92,Cas00}, where it was shown that the $A$-dependent
Feynman rules involving the $\Delta$ can be replaced by a set of
$A$-independent `reduced' vertices and propagator. 
In this framework, the reduced $\Delta$-propagator 
$\tilde G_{\alpha \beta}(p_\Delta)$ is given by \cite{Ami92}~: 
\begin{eqnarray}
\tilde G_{\alpha \beta}(p_\Delta) \,&=&\, 
{{\pdeldagger + M_\Delta} \over {p_\Delta^2 - M_\Delta^2}} \,
\Bigg \{ - g_{\alpha \beta} \,+\, {1 \over 3} \, \gamma_\alpha
\gamma_\beta \,+\, {1 \over {3 M_\Delta}} \, 
\left(\gamma_\alpha (p_\Delta)_\beta \,-\,
\gamma_\beta (p_\Delta)_\alpha \right) \,+\,
{2 \over {3 M_\Delta^2}} \, (p_\Delta)_\alpha \, (p_\Delta)_\beta
\,\Bigg\} \nonumber\\
&-&\, {2 \over {3 M_\Delta^2}} \, 
\Bigg \{ \gamma_\alpha (p_\Delta)_\beta \,+\, \gamma_\beta (p_\Delta)_\alpha   
\,-\, \gamma_\alpha \, (\pdeldagger - M_\Delta) \, \gamma_\beta \Bigg
\}, 
\label{eq:redprop}
\end{eqnarray}
where $\alpha$ and $\beta$ are the $\Delta$ four-vector indices, 
$p_\Delta$ is the four-momentum and $M_\Delta$ the mass of the $\Delta$. 
To take account of the finite width of the $\Delta$-resonance, we
follow the procedure of \cite{Ami92,Cas00} by using a complex pole
description for the $\Delta$-resonance excitation. This amounts to
the replacement~: 
\begin{equation}
M_\Delta \to M_\Delta \,-\, {i \over 2} \, \Gamma_\Delta \, ,
\label{eq:complexmass} 
\end{equation}
in the propagator of Eq.~(\ref{eq:redprop}). For the complex pole
parameter of the $\Delta(1232)$-resonance, 
we use the values given by the PDG \cite{PDG00}~: 
$M_\Delta$ = 1210 MeV, $\Gamma_\Delta$ = 100 MeV. 
Further on, when calculating the $\Delta$-resonance
contribution to the $\gamma p \to \gamma \pi^0 p$ process, 
we will discuss that this `complex mass scheme' guarantees electromagnetic 
gauge invariance, in contrast to using Breit-Wigner propagators with
energy-dependent widths. 
\newline
\indent
With the propagator of Eq.~(\ref{eq:redprop}), the amplitude for the 
$\Delta(1232)$-resonance contribution to the $\gamma p \to \pi^0 p$
process, corresponding to Fig.~\ref{fig:gap_piop_diag}(a), is then
given by~:  
\begin{eqnarray}
\varepsilon_\mu(k, \lambda) \; {\mathcal M}^{\mu}_{a} (\gamma p \to \pi^0 p)\, 
&=& \, i \, e \, {2 \over 3} \, 
{{f_{\pi N \Delta}} \over {m_\pi}} \, (p_\pi)^\alpha \, 
\varepsilon_\mu(k, \lambda) \nonumber\\
&\times& \bar N(p_N^{'}, s'_N) \, \tilde G_{\alpha \beta}(p_\Delta) \,
\left[ G_M \, \Gamma_M^{\beta \mu} \,+\, G_E \, \Gamma_E^{\beta \mu} \right] \,
N(p_N, s_N),
\label{eq:delpi0}
\end{eqnarray}
where $\varepsilon_\mu$ is the photon four-vector, 
$e$ is the proton electric charge 
($\alpha_{em}$ = $e^2 / (4 \pi)$ = 1/137) and 
$m_\pi$ the pion mass. The $\pi N \Delta$ coupling constant 
$f_{\pi N \Delta}$ in Eq.~(\ref{eq:delpi0}) is taken from the $\Delta
\to \pi N$ decay, which yields~: $f_{\pi N \Delta} \approx 1.95$.
Furthermore, in Eq.~(\ref{eq:delpi0}), 
$\Gamma_M^{\beta \mu}$ ($\Gamma_E^{\beta \mu}$) denote the
magnetic (electric) $\gamma N \Delta$ vertices respectively, and are
defined as~:
\begin{eqnarray}
\Gamma_M^{\beta \mu} \,&=&\, 
- {3 \over {2 M_N}} \, {1 \over {(M_N + M_\Delta)}}
\, \varepsilon^{\beta \mu \kappa \lambda} \, {1 \over 2} \, 
\left( p_\Delta + p_N\right)_\kappa \, k_\lambda \, \\
\Gamma_E^{\beta \mu} \,&=&\, - \Gamma_M^{\beta \mu}
\,-\, {6 \over {(M_\Delta + M_N)\,(M_\Delta - M_N)^2 \,M_N}} \nonumber\\
&\times&\, \Big(\,\varepsilon^{\beta \sigma \kappa \lambda} \, {1 \over 2} \, 
( p_\Delta + p_N )_\kappa \, k_\lambda \,\Big) \,
\Big(\,{{\varepsilon^{\mu}}_{\sigma}}^{\rho \tau} \, 
( p_\Delta )_\rho \, k_\tau \,\Big) \, i \, \gamma_5 \,.
\end{eqnarray} 
The magnetic $\gamma N \Delta$ coupling $G_M$ at the real photon point 
is taken from the MAID00 analysis~: $G_M(0)$ = 3.02 \cite{MAID}. 
The small electric $\gamma N \Delta$ coupling $G_E$, corresponding to
the $E2/M1$ ratio of about -2.5\% for the $\gamma N \to \Delta(1232)$
transition \cite{Beck00}, can be safely neglected for the sake of our
subsequent analysis. 
\newline
\indent
Starting from the $\Delta$-contribution of Eq.~(\ref{eq:delpi0}) to
the $\gamma p \to \pi^0 p$ process, we now construct the corresponding 
$\Delta$-contribution to the $\gamma p \to \gamma \pi^0 p$ process.
This is obtained by coupling a photon in a gauge invariant way to all
charged particles in Fig.~\ref{fig:gap_piop_diag}(a). 
This way we obtain in the first place the diagrams with a photon attached to
an external proton (Figs.~\ref{fig:gap_piogap_diag}(a1) and (a3)). 
Their contributions to the tensor ${\mathcal M}^{\nu \mu}$ of 
Eq.~(\ref{eq:cross3b}) is given by~: 
\begin{eqnarray}
{\mathcal M}^{\nu \mu}_{a1} (\gamma p \to \gamma \pi^0 p)\, 
&=& \, i \, e^2 \, {2 \over 3} \, 
{{f_{\pi N \Delta}} \over {m_\pi}} \, (p_\pi)^\alpha \, \nonumber\\
&\times& \bar N(p_N^{'}, s'_N) \, \tilde G_{\alpha \beta}(p_\Delta^{'}) \,
\left[ G_M \, \Gamma_M^{\beta \mu} \,+\, G_E \, \Gamma_E^{\beta \mu} \right] \,
\nonumber \\
&\times& \, {{\left( \pindagger - \kpdagger + M_N \right)} 
\over {-2 \, p_N \cdot k'}} \, 
\left[ \gamma^\nu - \kappa_p \, i \sigma^{\nu \rho} \, {{k'_\rho} \over {2 M_N}} \right] \, N(p_N, s_N),  \\
{\mathcal M}^{\nu \mu}_{a3} (\gamma p \to \gamma \pi^0 p)\, 
&=& \, i \, e^2 \, {2 \over 3} \, 
{{f_{\pi N \Delta}} \over {m_\pi}} \, (p_\pi)^\alpha \, \nonumber\\
&\times& \bar N(p_N^{'}, s'_N) \,
\left[ \gamma^\nu - \kappa_p \, i \sigma^{\nu \rho} \, {{k'_\rho} \over {2 M_N}} \right] 
\, {{\left( \poutdagger + \kpdagger + M_N \right)} 
\over {2 \, p_N^{'} \cdot k'}} \,\tilde G_{\alpha \beta}(p_\Delta) \,\nonumber\\
&\times& \,
\left[ G_M \, \Gamma_M^{\beta \mu} \,+\, G_E \, \Gamma_E^{\beta \mu} \right] 
\, N(p_N, s_N), 
\end{eqnarray}
where we also included, besides the electric coupling of the emitted photon
to the proton, the magnetic contribution ($\kappa_p = 1.79$ is the
proton anomalous magnetic moment). 
In the soft-photon limit ($k' \to 0$), the coupling of the photon to 
the external lines gives the only contribution. However, at finite
energy for the emitted photon, gauge invariance also requires  
the diagram with a photon attached to the intermediate
$\Delta^+$ (Fig.\ref{fig:gap_piogap_diag}(a2)), which corresponds to
the tensor~:
\begin{eqnarray} 
{\mathcal M}^{\nu \mu}_{a2} (\gamma p \to \gamma \pi^0 p)\, 
&=& \, - \,i \, e^2 \, {2 \over 3} \, 
{{f_{\pi N \Delta}} \over {m_\pi}} \, (p_\pi)^\alpha \, \nonumber\\
&\times& \bar N(p_N^{'}, s'_N) \, \tilde G_{\alpha \beta}(p_\Delta^{'}) \;
\Gamma_{\gamma \Delta \Delta}^{\, \nu \beta \beta'} 
\; \tilde G_{\beta' \delta}(p_\Delta)
\left[ G_M \,\Gamma_M^{\delta \mu} \,+\, G_E \,\Gamma_E^{\delta \mu} \right] \,
N(p_N, s_N),
\end{eqnarray}
where $\Gamma_{\gamma \Delta \Delta}$ is the $\gamma \Delta \Delta$
vertex~: 
\begin{equation}
\Gamma_{\gamma \Delta \Delta}^{\, \nu \beta \beta'} \,=\, 
g^{\beta \beta'} \, \left( \gamma^\nu \,-\, i \, \kappa_\Delta \, 
\sigma^{\nu \lambda} \, {{k'_\lambda} \over {2 M_\Delta}} \right)
\,+\, {1 \over 3} \, \left( \gamma^\beta \gamma^\nu \gamma^{\beta'}
  \,-\, \gamma^\beta \, g^{\nu \beta'} \,-\, \gamma^{\beta'} \, g^{\nu \beta} 
\right),
\label{eq:gadeldel}
\end{equation}
containing the photon coupling to a spin 3/2 field 
of a point-like particle (i.e. $\kappa_\Delta = 0$) \cite{Ami92}, 
as well as the $\Delta$ anomalous magnetic dipole moment contribution,
which is proportional to $\kappa_\Delta$. 
\newline
\indent
The vertex for a point- spin-3/2 particle follows from the requirement
of gauge invariance. Indeed, only the sum of the 3 diagrams 
Fig.~\ref{fig:gap_piogap_diag}(a1,a2,a3) is gauge-invariant, which is
expressed through the electromagnetic Ward identity relating the
$\gamma \Delta \Delta$ vertex and the $\Delta$-reduced propagator~:
\begin{equation}
\left( p_{\Delta}^{'} - p_\Delta \right)_\nu \, 
\tilde G_{\alpha \kappa}(p_{\Delta}^{'}) \; 
\Gamma_{\gamma \Delta \Delta}^{\, \nu \kappa \lambda} \;
\tilde G_{\lambda \beta}(p_{\Delta}) \, 
=\, - \, \tilde G_{\alpha \beta}(p_{\Delta})  
\, + \, \tilde G_{\alpha \beta}(p_{\Delta}^{'}) \, . 
\label{eq:wardiden}
\end{equation}
\indent
To include the finite width of the $\Delta$, the `complex mass scheme'
was advocated in Ref.~\cite{Ami92}, 
i.e. the replacement of $M_\Delta$ by the complex
mass of Eq.~(\ref{eq:complexmass}) {\it everywhere} in the propagator
Eq.~(\ref{eq:redprop}), and not only in its denominator. In doing so,
it is easily checked that the Ward identity of Eq.~(\ref{eq:wardiden})
still holds. On the other hand, when using Breit-Wigner propagators,
replacing $M_\Delta^2$ in the denominator of Eq.~(\ref{eq:redprop}) by
$M_\Delta^2 \,-\, i \, M_\Delta \,\Gamma_\Delta(W)$, and with 
$\Gamma_\Delta(W)$ an energy dependent width, the $\Delta$
propagators before and after the emission of the photon have different
widths (except in the soft-photon limit $k' \to 0$, 
where the contribution of diagram
Fig.~\ref{fig:gap_piogap_diag}(a2) vanishes relative to those of 
Fig.~\ref{fig:gap_piogap_diag}(a1) and (a3)). 
Therefore, when using energy-dependent widths 
at finite energy of the emitted photon, it is not possible
to maintain the Ward identity of Eq.~(\ref{eq:wardiden}) with the
point vertex ($\kappa_\Delta = 0$) of Eq.~(\ref{eq:gadeldel}).
\newline
\indent
In recent years, the issue of a gauge invariant definition of resonant
masses, widths and partial widths has newly received attention in view
of high precision measurements of the properties of the $W$ and $Z$
gauge bosons at LEP \cite{Sir91,Stu93,Vel94,Lop95}. In particular, it has
been shown in Refs.~\cite{Sir91,Stu93,Vel94,Lop95} that a complex pole
definition of a resonance provides a systematic way to maintain gauge
invariance of the amplitude at any order of perturbation theory. 
In particular, one-loop corrections to $W^{\pm}$ 
propagators and vertices have been calculated  
through a fermion loop \cite{Baur95,Beu97}, 
and it has been shown in the limit of massless fermions (leptons,
quarks) in the loop that the simplest prescription to 
obtain gauge invariant amplitudes in the presence of finite width
effects is to perform a Laurent expansion of the full amplitude around
the complex pole position of the resonance \cite{Lop95}. 
A similar observation has been made in Ref.~\cite{Lop00}, 
where the $\rho^{\pm}$ vector meson contribution to the 
$\tau \to \pi \pi \nu \gamma$ decay has been studied as a source to
extract the magnetic dipole moment of the $\rho$ meson. 
The absorptive one-loop corrections to the propagator
and vertices of the charged $\rho$ vector meson were calculated in
Ref.~\cite{Lop00} through a pion loop.  
In the limit of vanishing mass of the pions in the loop, 
it was further shown that the `complex mass scheme', 
replacing the $\rho$ mass by a complex
parameter in the lowest order expressions for propagators and vertices,
provides a gauge invariant description in the presence of the finite
width of the $\rho$ meson. 
\newline
\indent
This `complex mass scheme' prescription was then applied in 
Ref.~\cite{Cas00} to the propagator and vertices of the $\Delta^{++}$ 
resonance, in the description of the $\pi^+ p \to \gamma \pi^+ p$
reaction. It was argued for the $\Delta$ resonance, 
similar to the spin-1 case, that the constraint imposed by gauge invariance
leads to this complex mass scheme. An explicit calculation of the
absorption corrections to $\Delta$ propagator and vertices through 
$\pi N$ loops would be very valuable to check this conjecture. In this
work, we will follow the `complex mass scheme' prescription
of Eq.~(\ref{eq:complexmass}), and apply it to the description of the
$\Delta^+$ resonance in the $\gamma p \to \gamma \pi^0 p$
process. This guarantees electromagnetic gauge invariance as expressed
in Eq.~(\ref{eq:wardiden}).
\newline
\indent
Besides the vertex for a point particle of spin 3/2, which is constrained
by gauge invariance, we also included in Eq.~(\ref{eq:gadeldel}) the
contribution of the $\Delta$ anomalous magnetic dipole moment, proportional to
$\kappa_\Delta$. This magnetic moment contribution is gauge invariant
by itself. The extraction of the value of $\kappa_\Delta$ is the
principal motivation for constructing the present model to analyse the
$\gamma p \to \gamma \pi^0 p$ reaction. 
The $\Delta^+$ magnetic moment $\mu_{\Delta +}$ 
is then obtained from the $\Delta^+$ anomalous magnetic moment
$\kappa_{\Delta +}$ as~:
\begin{equation} 
\mu_{\Delta +} \equiv ( 1 + \kappa_{\Delta +}) \, 
{e \over {2 M_\Delta}} \,=\, 
( 1 + \kappa_{\Delta +}) \, {{M_N} \over {M_\Delta}} \, \mu_N \; ,
\end{equation} where 
$\mu_N \equiv e /(2 M_N)$ is the usual nuclear magneton. The SU(6)
value predicts the value $\kappa_{\Delta +} = \kappa_p = 1.79$.
Assuming SU(2) isospin symmetry, the PDG range for $\mu_{\Delta^{++}}$
corresponds for $\Delta^+$ to the range 
$\mu_{\Delta^+}$ = 1.85 - 3.75 $\mu_N$, which translates to 
the range $\kappa_{\Delta^+}$ = 1.43 - 3.92.
\newline
\indent
In addition to the magnetic dipole moment (M1), the $\Delta(1232)$
also has static monopole and quadrupole Coulomb moments (C0, C2) and a
magnetic octupole moment (M3). It is obvious that the Coulomb moment
does not couple to the emitted photon, and the magnetic octupole
moment is suppressed by two additional powers in the photon momentum.
There is, in principle, a contribution from electric (transverse)
quadrupole radiation (E2). However this is strongly suppressed by the
fact that E2 radiation vanishes by time-reversal invariance if the
initial and final states are exactly the same. In fact, such radiation
can only contribute by off-shell and recoil effects of the
intermediate resonance. In view of the small ratio $R = E2/M1 \approx
- 2.5 \%$ for the $N \Delta$ transition, it is therefore
(unfortunately !) very unlikely that one will ever learn about the
quadrupole moment of the $\Delta(1232)$ by radiative pion
photoproduction.
In the present description, we therefore keep only the magnetic dipole
moment of the $\Delta(1232)$.

\subsection{Vector meson ($\omega$) exchange processes}

Although the $\gamma p \to \pi^0 p$ process in the $\Delta(1232)$
resonance region is dominated by the $\Delta$-excitation process, it
is well known that an accurate description also necessitates the
inclusion of non-resonant mechanisms. In the following, we discuss 
such non-resonant mechanisms and calculate them for the $\gamma p \to
\gamma \pi^0 p$ process. 
\newline
\indent
At the high energy side of the $\Delta$(1232)-region, 
the prominent non-resonant contribution to the 
$\gamma p \to \pi^0 p$ reaction comes from vector-meson exchange,
dominantly $\omega$-exchange as shown in Fig.~\ref{fig:gap_piop_diag}(b). 
The amplitude of the $\omega$-exchange contribution is given by~:
\begin{eqnarray} 
\varepsilon_\mu(k, \lambda) \; 
{\mathcal M}^{\mu}_{b} (\gamma p \to \pi^0 p)\, &=&\,
- i e \, g_{\omega NN} \, {{g_{\omega \pi \gamma}} \over {m_\pi}} \,
{1 \over {t - m_\omega^2}} \, 
\varepsilon^{\lambda \mu \alpha \beta} \, k_\lambda \, 
(k - p_\pi)_\alpha \, \varepsilon_\mu(k, \lambda) \nonumber\\
&\times& \, \bar N(p_N^{'}, s'_N) \, 
\left[ \gamma_\beta + \kappa_\omega \, i \sigma_{\beta \sigma} \, 
{{(k - p_\pi)^\sigma} \over {2 M_N}} \right] \, N(p_N, s_N), 
\end{eqnarray}
where $m_\omega = 0.7826$ GeV is the $\omega$-meson mass. 
The $\omega \pi \gamma$ coupling $g_{\omega \pi \gamma}$ is well known
from the radiative decay $\omega \to \pi \gamma$, which leads to the
value~: $g_{\omega \pi \gamma}$ = 0.314.
The $\omega NN$ coupling $g_{\omega NN}$ is taken from a high-energy
analysis of $\gamma \, p \, \to \, \pi^0 \, p$ ~: $g_{\omega NN}$ = 15 
\cite{GLV}. The (isoscalar) tensor coupling $\kappa_\omega$ is
generally found to be small and is set equal to zero
in the present analysis, i.e. $\kappa_\omega = 0$. 
\newline
\indent
When coupling a photon in a gaugeinvariant way to both charged proton
lines in Fig.~\ref{fig:gap_piop_diag}(b), one obtains the
corresponding diagrams of Fig.~\ref{fig:gap_piogap_diag}(b1) and (b2)
for the $\gamma p \to \gamma \pi^0 p$ reaction. 
Their contributions to the tensor ${\mathcal M}^{\nu \mu}$ of 
Eq.~(\ref{eq:cross3b}) are given by~: 
\begin{eqnarray} 
{\mathcal M}^{\nu \mu}_{b1} (\gamma p \to \gamma \pi^0 p)\,&=&\, 
- i e^2 \, g_{\omega NN} \, {{g_{\omega \pi \gamma}} \over {m_\pi}} \,
{1 \over {t - m_\omega^2}} \, 
\varepsilon^{\lambda \mu \alpha \beta} \, k_\lambda \, 
(k - p_\pi)_\alpha \, \nonumber\\
&\times& \,
\bar N(p_N^{'}, s'_N) \, 
\left[ \gamma_\beta + \kappa_\omega \, i \sigma_{\beta \sigma} \, 
{{(k - p_\pi)^\sigma} \over {2 M_N}} \right] \,\nonumber\\
&\times&\, {{\left( \pindagger - \kpdagger + M_N \right)} 
\over {-2 \, p_N \cdot k'}} \, 
\left[ \gamma^\nu - \kappa_p \, i \sigma^{\nu \rho} \, {{k'_\rho} \over {2 M_N}} \right] 
\, N(p_N, s_N), \\
{\mathcal M}^{\nu \mu}_{b2} (\gamma p \to \gamma \pi^0 p)\,&=&\, 
- i e^2 \, g_{\omega NN} \, {{g_{\omega \pi \gamma}} \over {m_\pi}} \,
{1 \over {t - m_\omega^2}} \, 
\varepsilon^{\lambda \mu \alpha \beta} \, k_\lambda \, 
(k - p_\pi)_\alpha \, \nonumber\\
&\times& \,
\bar N(p_N^{'}, s'_N) \, 
\left[ \gamma^\nu - \kappa_p \, i \sigma^{\nu \rho} \, {{k'_\rho} \over {2 M_N}} \right] \,
{{\left( \poutdagger + \kpdagger + M_N \right)} 
\over {2 \, p_N^{'} \cdot k'}} \, \nonumber\\
&\times& \,  
\left[ \gamma_\beta + \kappa_\omega \, i \sigma_{\beta \sigma} \, 
{{(k - p_\pi)^\sigma} \over {2 M_N}} \right] \, N(p_N, s_N), 
\end{eqnarray}

\subsection{Born term contributions}

At the low energy side of the $\Delta(1232)$-region, the prominent
non-resonant contribution to the $\gamma p \to \pi^0 p$ reaction comes
from the Born contributions as shown in 
Fig.~\ref{fig:gap_piop_diag}(c1) and (c2).
They are evaluated here using pseudo-vector $\pi NN$ coupling, which
yields the amplitudes~:  
\begin{eqnarray}
\varepsilon_\mu(k, \lambda) \; 
{\mathcal M}^{\mu}_{c1} (\gamma p \to \pi^0 p)\, 
&=& - \, e \, {{f_{\pi N N}} \over {m_\pi}} \; \varepsilon_\mu(k, \lambda) \;
\bar N(p_N^{'}, s'_N) \, \ppidagger \, \gamma_5 \;
{{\left( \pindagger + \kdagger + M_N \right)} \over {2 \, p_N \cdot k}}
 \nonumber\\
&\times&\, \left[ \gamma^\mu + \kappa_p \, i \sigma^{\mu \rho} \, 
{{k_\rho} \over {2 M_N}} \right] \, 
N(p_N, s_N), \\
\varepsilon_\mu(k, \lambda) \; 
{\mathcal M}^{\mu}_{c2} (\gamma p \to \pi^0 p)\, 
&=& - \, e \, {{f_{\pi N N}} \over {m_\pi}} \; \varepsilon_\mu(k, \lambda) \;
\bar N(p_N^{'}, s'_N) 
\, \left[ \gamma^\mu + \kappa_p \, i \sigma^{\mu \rho} \, {{k_\rho} \over
    {2 M_N}} \right] \nonumber\\ 
&\times& \, 
{{\left( \poutdagger - \kdagger + M_N \right)} \over {-2 \, p_N^{'} \cdot k}} 
\ppidagger \, \gamma_5 \, N(p_N, s_N) \, . 
\end{eqnarray}
The $\pi NN$ coupling constant is given by its well known value~:
$f_{\pi NN}^2/(4 \pi) = 0.08$ \cite{vpi99}. 
\newline
\indent
The corresponding contributions to the $\gamma p \to \gamma \pi^0 p$
reaction are obtained by coupling a photon in a gauge invariant way to
all proton lines in Fig.~\ref{fig:gap_piop_diag}(c1) and (c2). 
This yields the six diagrams of Fig.~\ref{fig:gap_piogap_diag}(c1-c6)
whose contributions to the tensor ${\mathcal M}^{\nu \mu}$ of 
Eq.~(\ref{eq:cross3b}) are given by~: 
\begin{eqnarray}
&&{\mathcal M}^{\nu \mu}_{c1} (\gamma p \to \gamma \pi^0 p)\, 
= - \, e^2 \, {{f_{\pi N N}} \over {m_\pi}} \; 
\bar N(p_N^{'}, s'_N) \, \ppidagger \, \gamma_5 \;
{{\left( \pindagger + \kdagger - \kpdagger + M_N \right)} 
\over {2 \, p_N \cdot (k - k') \,-\, 2 \, k \cdot k' }} \nonumber\\
&&\hspace{2cm} \times\, \left[ \gamma^\mu + \kappa_p \, i \sigma^{\mu \rho} \, 
{{k_\rho} \over {2 M_N}} \right] \, 
{{\left( \pindagger - \kpdagger + M_N \right)} 
\over {-2 \, p_N \cdot k'}} \, 
\left[ \gamma^\nu - \kappa_p \, i \sigma^{\nu \sigma} \, 
{{k'_\sigma} \over {2 M_N}} \right] \,
N(p_N, s_N), 
\label{eq:born1} \\
&&{\mathcal M}^{\nu \mu}_{c2} (\gamma p \to \gamma \pi^0 p)\, 
= - \, e^2 \, {{f_{\pi N N}} \over {m_\pi}} \; 
\bar N(p_N^{'}, s'_N) \,\ppidagger \, \gamma_5 \;
{{\left( \pindagger + \kdagger - \kpdagger + M_N \right)} 
\over {2 \, p_N \cdot (k - k') \,-\, 2 \, k \cdot k' }} \nonumber\\
&&\hspace{2cm} \times \, 
\left[ \gamma^\nu - \kappa_p \, i \sigma^{\nu \sigma} \, 
{{k'_\sigma} \over {2 M_N}} \right] \,
{{\left( \pindagger + \kdagger + M_N \right)} \over {2 \, p_N \cdot k}} \, 
\left[ \gamma^\mu + \kappa_p \, i \sigma^{\mu \rho} \, 
{{k_\rho} \over {2 M_N}} \right] \, 
N(p_N, s_N), 
\label{eq:born2} \\ 
&&{\mathcal M}^{\nu \mu}_{c3} (\gamma p \to \gamma \pi^0 p)\, 
= - \, e^2 \, {{f_{\pi N N}} \over {m_\pi}} \; 
\bar N(p_N^{'}, s'_N) \,
\left[ \gamma^\nu - \kappa_p \, i \sigma^{\nu \sigma} \, 
{{k'_\sigma} \over {2 M_N}} \right] \,
{{\left( \poutdagger + \kpdagger + M_N \right)} \over 
{2 \, p_N^{'} \cdot k'}} \, \nonumber\\
&&\hspace{2cm} \times \, 
\ppidagger \, \gamma_5 \;
{{\left( \pindagger + \kdagger + M_N \right)} \over {2 \, p_N \cdot k}} \,
\left[ \gamma^\mu + \kappa_p \, i \sigma^{\mu \rho} \, 
{{k_\rho} \over {2 M_N}} \right] \, 
N(p_N, s_N), 
\label{eq:born3} \\
&&{\mathcal M}^{\nu \mu}_{c4} (\gamma p \to \gamma \pi^0 p)\, 
= - \, e^2 \, {{f_{\pi N N}} \over {m_\pi}} \; 
\bar N(p_N^{'}, s'_N) \, 
\left[ \gamma^\mu + \kappa_p \, i \sigma^{\mu \rho} \, 
{{k_\rho} \over {2 M_N}} \right] \, 
{{\left( \poutdagger - \kdagger + M_N \right)} 
\over {-2 \, p_N^{'} \cdot k}} \nonumber\\
&&\hspace{2cm} \times\, 
\ppidagger \, \gamma_5 \;
{{\left( \pindagger - \kpdagger + M_N \right)} 
\over {-2 \, p_N \cdot k'}} \, 
\left[ \gamma^\nu - \kappa_p \, i \sigma^{\nu \sigma} \, 
{{k'_\sigma} \over {2 M_N}} \right] \,
N(p_N, s_N), 
\label{eq:born4} \\
&&{\mathcal M}^{\nu \mu}_{c5} (\gamma p \to \gamma \pi^0 p)\, 
= - \, e^2 \, {{f_{\pi N N}} \over {m_\pi}} \; 
\bar N(p_N^{'}, s'_N) \,
\left[ \gamma^\mu + \kappa_p \, i \sigma^{\mu \rho} \, 
{{k_\rho} \over {2 M_N}} \right] \, 
{{\left( \poutdagger - \kdagger + M_N \right)} 
\over {-2 \, p_N^{'} \cdot k}} \nonumber\\
&&\hspace{2cm} \times \, 
\left[ \gamma^\nu - \kappa_p \, i \sigma^{\nu \sigma} \, 
{{k'_\sigma} \over {2 M_N}} \right] \,
{{\left( \poutdagger + \kpdagger - \kdagger + M_N \right)} 
\over {-2 \, p_N^{'} \cdot (k - k') \,-\, 2 \, k \cdot k' }} \,
\ppidagger \, \gamma_5 \, N(p_N, s_N), 
\label{eq:born5} \\ 
&&{\mathcal M}^{\nu \mu}_{c6} (\gamma p \to \gamma \pi^0 p)\, 
= - \, e^2 \, {{f_{\pi N N}} \over {m_\pi}} \; 
\bar N(p_N^{'}, s'_N) \,
\left[ \gamma^\nu - \kappa_p \, i \sigma^{\nu \sigma} \, 
{{k'_\sigma} \over {2 M_N}} \right] \,
{{\left( \poutdagger + \kpdagger + M_N \right)} \over 
{2 \, p_N^{'} \cdot k'}} \, \nonumber\\
&&\hspace{2cm} \times \, 
\left[ \gamma^\mu + \kappa_p \, i \sigma^{\mu \rho} \, 
{{k_\rho} \over {2 M_N}} \right] \, 
{{\left( \poutdagger - \kdagger + \kpdagger + M_N \right)} \over 
{-2 \, p_N^{'} \cdot (k - k') \,-\, 2 \, k \cdot k'}} \,
\ppidagger \, \gamma_5 \, N(p_N, s_N) \, . 
\label{eq:born6}
\end{eqnarray}
Note that all six amplitudes of Eqs.(\ref{eq:born1}-\ref{eq:born6})
are needed to satisfy gauge invariance with respect to both the
incoming and outgoing photons.

\section{Results for $\gamma \, p \to  \gamma \, \pi^0 \, p$ observables and discussion}
\label{results}

\begin{figure}[ht]
\epsfxsize=13 cm
\centerline{\epsffile{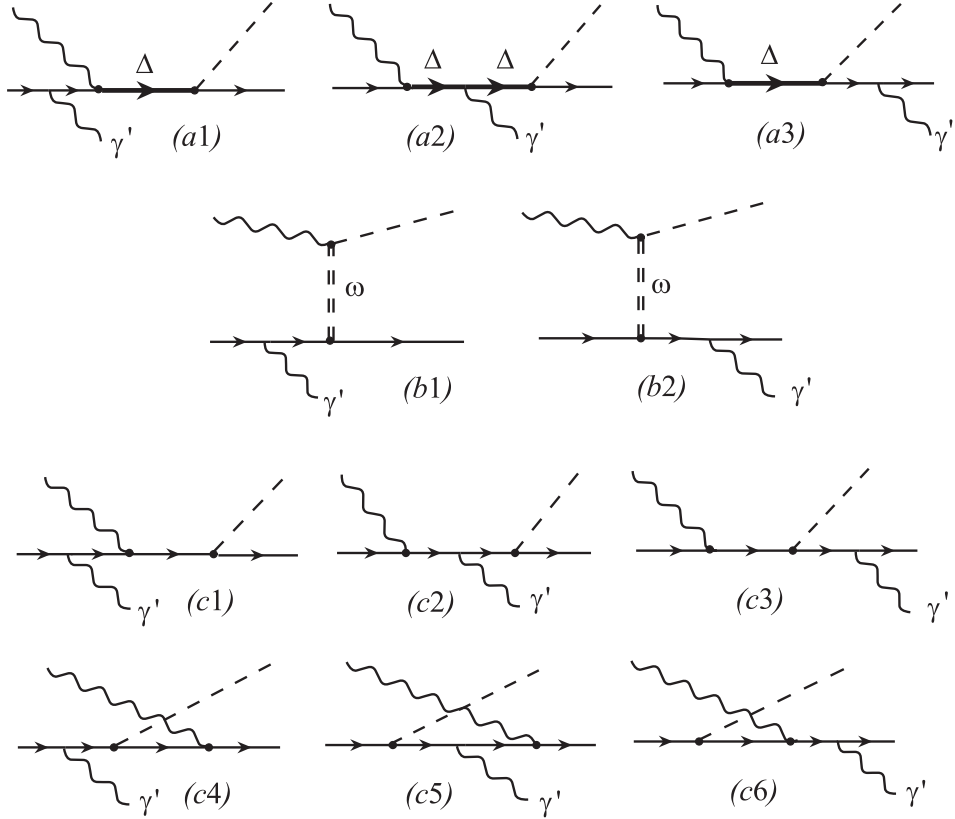}}
\vspace{.5cm}
\caption[]{Diagrams considered in the calculation of 
the $\gamma \, p \, \to \, \gamma^{\, '} \, \pi^0 \, p$ reaction in the
$\Delta(1232)$ region, obtained by gauge invariant coupling of a photon 
to the diagrams of Fig.~\ref{fig:gap_piop_diag}. 
This yields the $\Delta$-resonance diagrams (a1-a3),
$\omega$-exchange diagrams (b1-b2), and Born diagrams where the photon
is emitted from all proton lines (c1-c6). 
Note that, with respect to the outgoing
photon, the sum of the three diagrams (a1-a3) is gauge invariant, 
and the sum of (b1-b2) is gauge invariant. 
Furthermore, for the Born terms, 
the six diagrams (c1-c6) are required to satisfy gauge invariance with
respect to both the incoming and outgoing photons. 
In comparison with the diagrams of Fig.~\ref{fig:gap_piop_diag} for 
$\gamma \, p \, \to \, \pi^0 \, p$, there
is only {\it one} new parameter which enters here at the $\gamma \Delta
\Delta$ vertex (a2), namely the $\Delta^+$ anomalous magnetic moment
$\kappa_{\Delta +}$. }
\label{fig:gap_piogap_diag}
\end{figure}

\begin{figure}[ht]
\epsfxsize=11 cm
\centerline{\epsffile{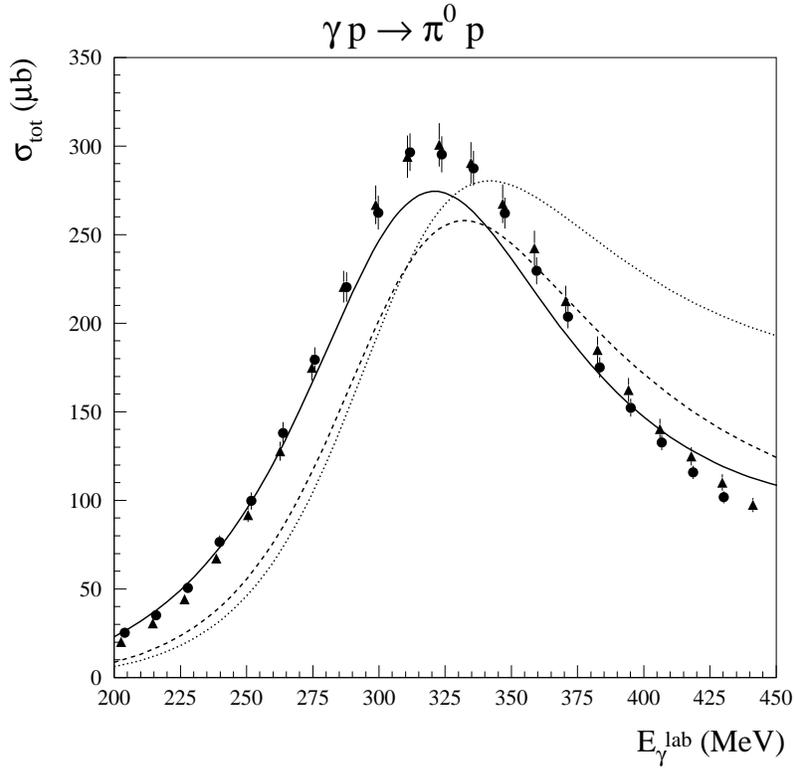}}
\caption[]{Total cross section for 
the $\gamma \, p \, \to \, \pi^0 \, p$ reaction in the $\Delta(1232)$-region. 
The calculations correspond to the mechanisms of
Fig.~\ref{fig:gap_piop_diag} : $\Delta$-excitation (dotted curve), 
$\Delta$ + $\omega$-exchange (dashed curve), 
and the sum of the three mechanisms (full curves). 
The difference between the full curves and the data is 
understood as originating mainly from $\pi N$ rescattering
mechanisms. The MAMI data are from Ref.~\cite{piotot_macc} (circles) and
from Ref.~\cite{piotot_ahr} (triangles).}
\label{fig:tot_ppio}
\end{figure}

\begin{figure}[ht]
\epsfxsize=10 cm
\centerline{\epsffile{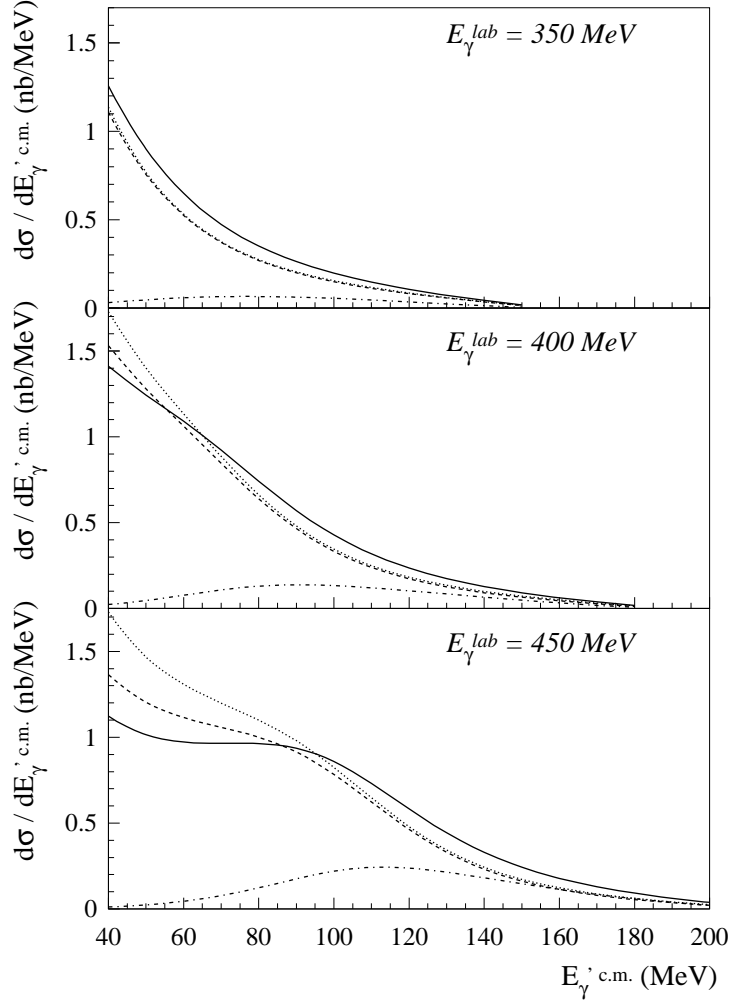}}
\caption[]{Outgoing photon c.m. energy dependence of the cross section $d
  \sigma / d E_\gamma^{\, ' c.m.}$ (integrated over photon and pion angles) 
of the $\gamma \, p \, \to \, \gamma \, \pi^0 \, p$ reaction. 
The calculation with only the $\Delta$ anomalous magnetic moment
contribution (a2) of Fig.~\ref{fig:gap_piogap_diag} is 
shown by the dashed-dotted curves. 
The calculation of the $\Delta$-resonance mechanisms
(a1-a3) of Fig.~\ref{fig:gap_piogap_diag} is shown by the dotted curves. 
The sum of $\Delta$-resonance and $\omega$-exchange mechanisms
(diagrams (b1-b2) of Fig.~\ref{fig:gap_piogap_diag}) is given by the
dashed curves, and the sum of all mechanism of
Fig.~\ref{fig:gap_piogap_diag} (including the Born terms (c1-c6)) is
given by the full curves. The value of the $\Delta^+$ anomalous magnetic
moment was set equal to $\kappa_{\Delta +}$ = 3 .}
\label{fig:gap_ppioga_en1}
\end{figure}

\begin{figure}[ht]
\epsfxsize=10 cm
\centerline{\epsffile{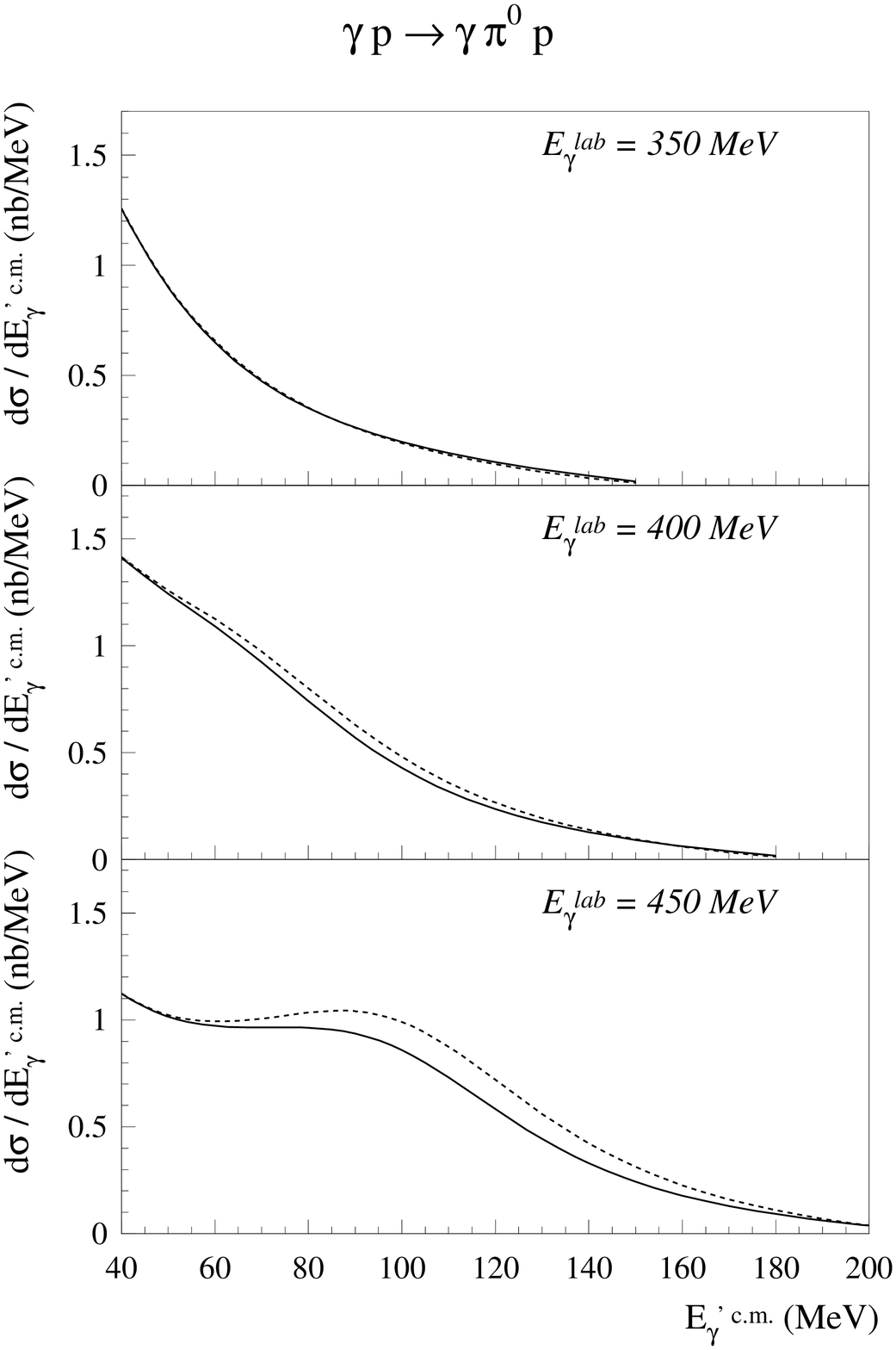}}
\caption[]{Outgoing photon c.m. energy dependence of the cross section $d
  \sigma / d E_\gamma^{\, ' c.m.}$ (integrated over photon and 
pion angles) of the $\gamma \, p \, \to \, \gamma \, \pi^0 \, p$
reaction. The calculations are the total results (including all
diagrams of Fig.~\ref{fig:gap_piogap_diag}) for the values~:
$\kappa_{\Delta +}$ = 0 (dashed curves) 
and $\kappa_{\Delta +}$ = 3 (full curves).}
\label{fig:gap_ppioga_en2}
\end{figure}

\begin{figure}[ht]
\epsfxsize=10 cm
\centerline{\epsffile{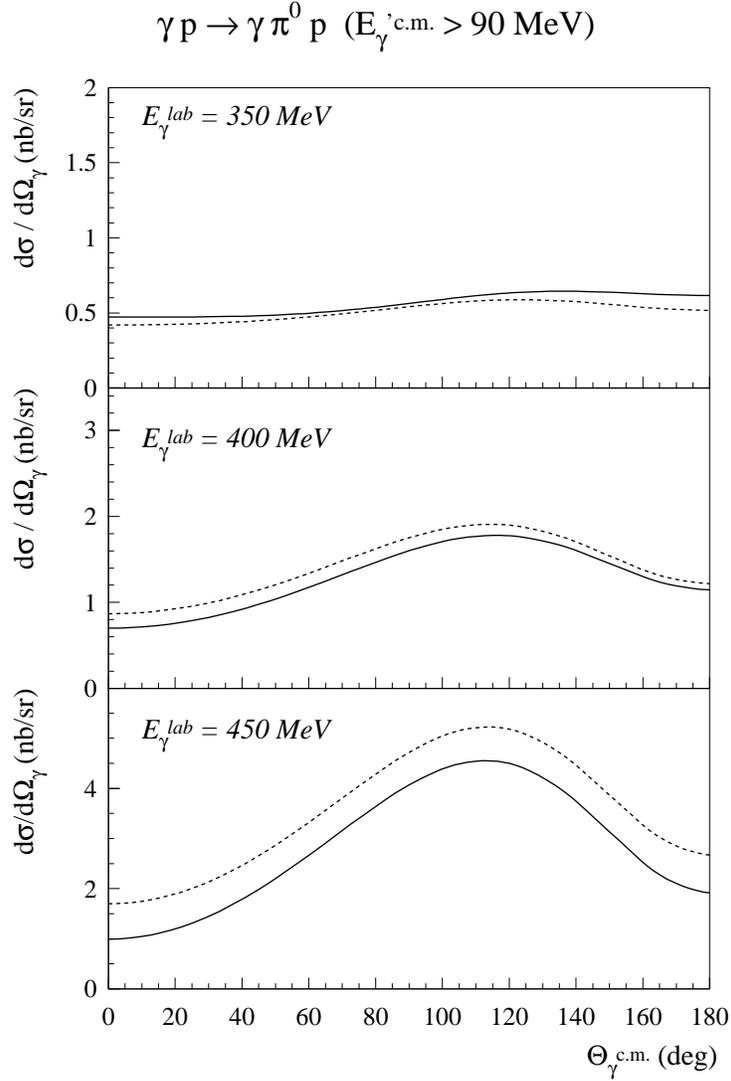}}
\caption[]{Outgoing photon c.m. angular dependence of the cross section $d
  \sigma / d \Omega_\gamma^{\, c.m.}$ 
of the $\gamma \, p \, \to \, \gamma \, \pi^0 \, p$ reaction. 
The cross section is integrated over the pion angles and over 
the outgoing photon energy range $E_\gamma^{\, ' c.m.} > 90$~MeV. 
The calculations are the total results (including all
diagrams of Fig.~\ref{fig:gap_piogap_diag}) for the values~:
$\kappa_{\Delta +}$ = 0 (dashed curves) 
and $\kappa_{\Delta +}$ = 3 (full curves).} 
\label{fig:gap_ppioga_ang_taps}
\end{figure}

\begin{figure}[ht]
\epsfxsize=10 cm
\centerline{\epsffile{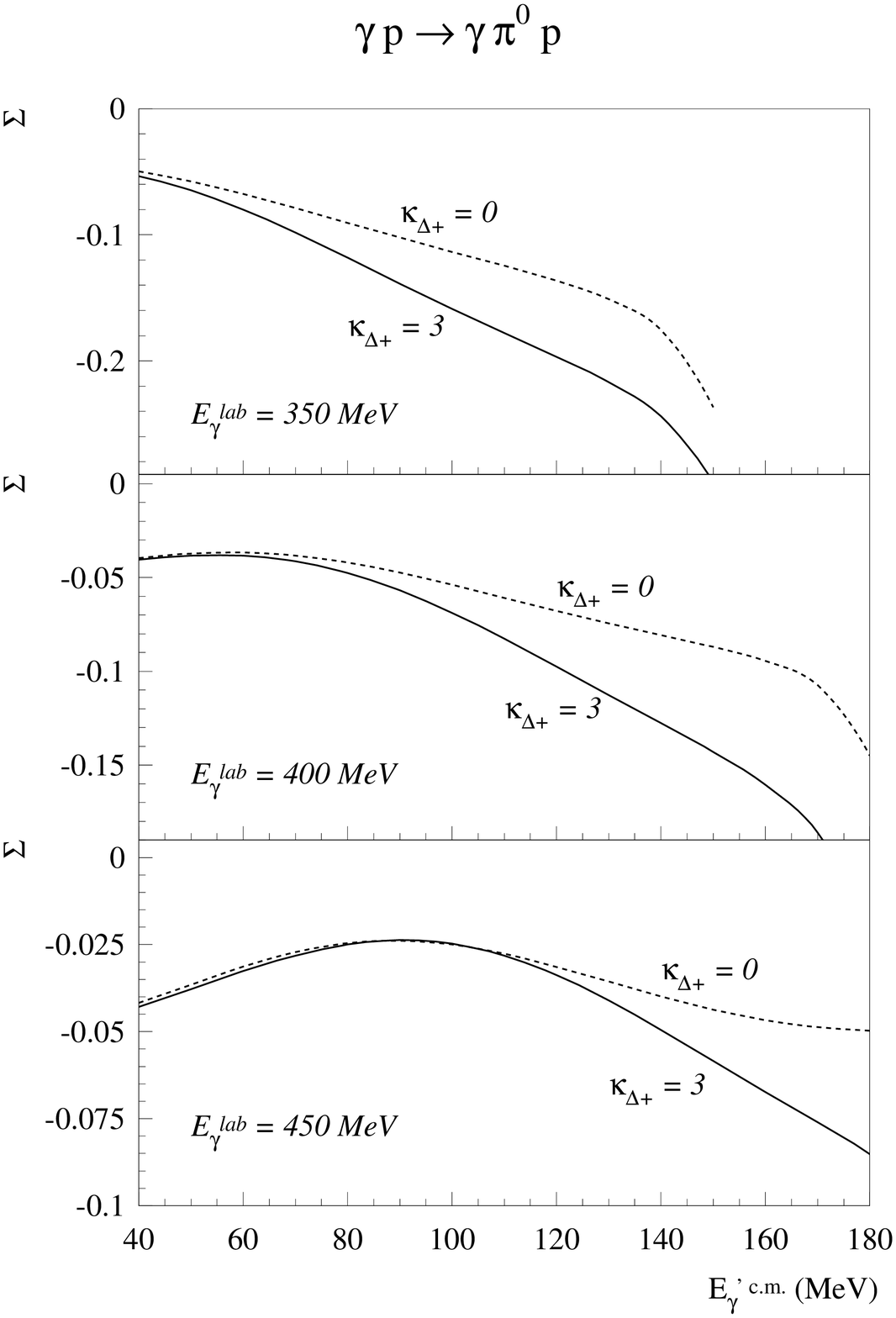}}
\caption[]{Outgoing photon c.m. energy dependence of the 
$\gamma \, p \, \to \, \gamma \, \pi^0 \, p$ 
photon asymmetry $\Sigma$, integrated over the photon and 
pion angles, and differential w.r.t. the outgoing photon energy 
$E_\gamma^{\, ' c.m.}$,
as in Fig.~\ref{fig:gap_ppioga_en2}. 
The calculations are the total results (including all
diagrams of Fig.~\ref{fig:gap_piogap_diag}), for the two values of
  $\kappa_{\Delta +}$ as indicated on the curves.}
\label{fig:gap_ppioga_en_asymm}
\end{figure}

\begin{figure}[ht]
\epsfxsize=13 cm
\centerline{\epsffile{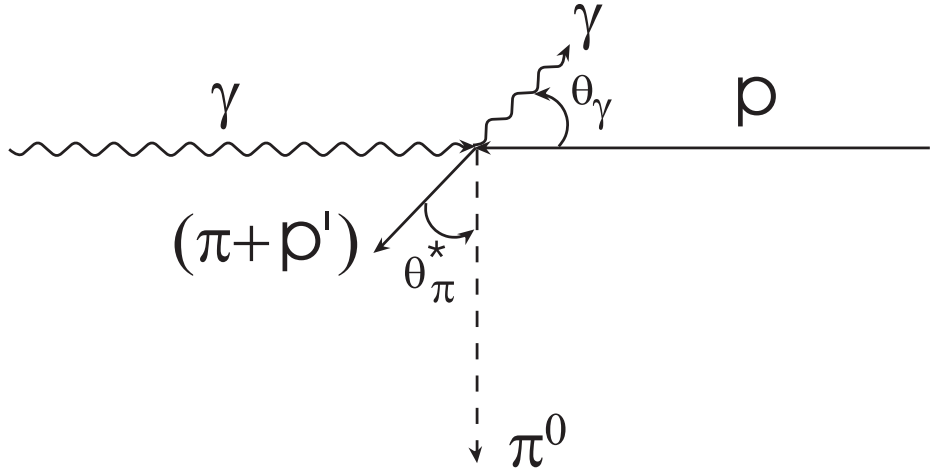}}
\caption[]{Definition of angles of the $\gamma \, p \, \to \, \gamma
  \, \pi^0 \, p$ reaction : $\theta_\gamma$ is the photon angle in the 
c.m. system, $\theta_\pi^{*}$ is the pion angle in the $(\pi^0 p)$ rest
frame w.r.t. the direction of the $(\pi^0 p)$ c.m. momentum. 
The pion momentum displayed in the figure corresponds to an out-of-plane angle 
$\Phi_\pi^* \,=\,0^o$.}
\label{fig:kinem_forw}
\end{figure}

\begin{figure}[ht]
\epsfxsize=11 cm
\centerline{\epsffile{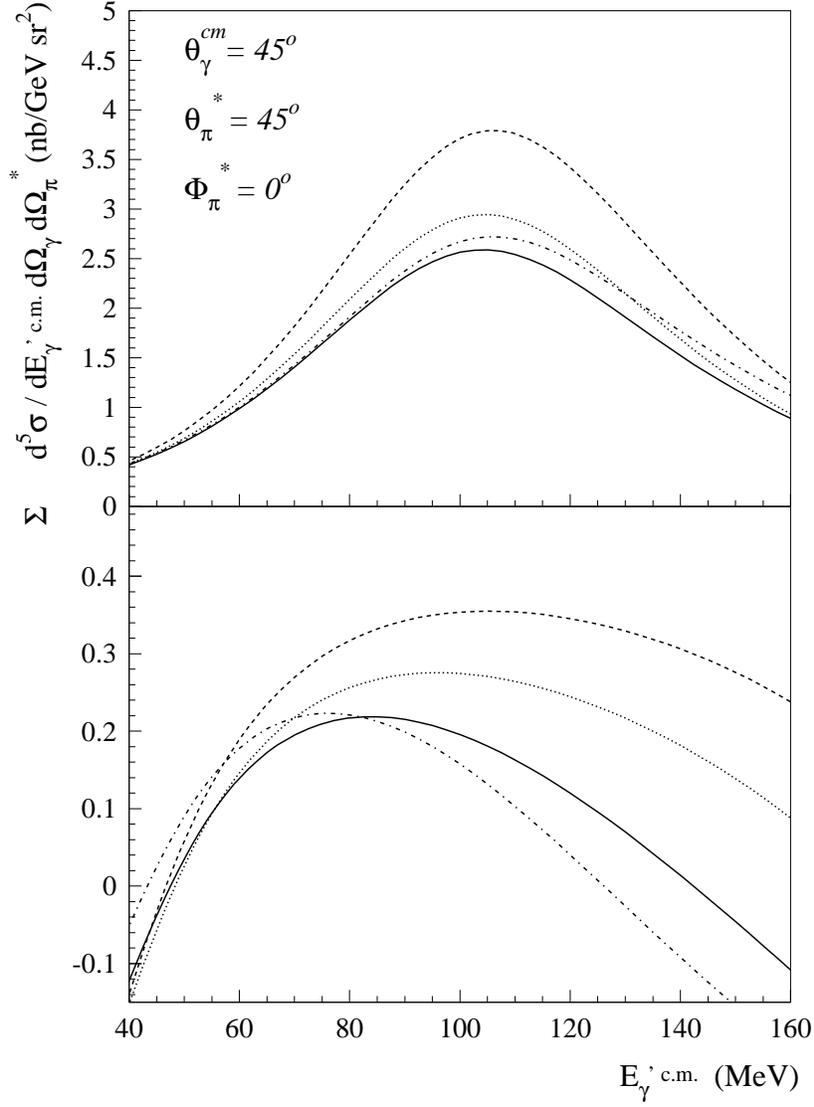}}
\caption[]{Five-fold differential cross section $d^5 \sigma$ (upper
  panel), and corresponding photon asymmetry $\Sigma$ (lower panel)
  for the $\gamma \, p \, \to \, \gamma \, \pi^0 \, p$ reaction as
  function of the outgoing photon c.m. energy.
The angles are defined as in Fig.~\ref{fig:kinem_forw}, and 
the kinematics considered here corresponds 
 (approximately) to the situation depicted in that figure. 
The calculations are the total results (including all
diagrams of Fig.~\ref{fig:gap_piogap_diag}), for the values ~:
$\kappa_{\Delta +}$ = 0 (dashed curves), 
$\kappa_{\Delta +}$ = 1.5 (dotted curves), 
$\kappa_{\Delta +}$ = 3 (full curves), 
$\kappa_{\Delta +}$ = 4.5 (dashed-dotted curves).}
\label{fig:diffasymm4}
\end{figure}

\begin{figure}[ht]
\epsfxsize=13 cm
\centerline{\epsffile{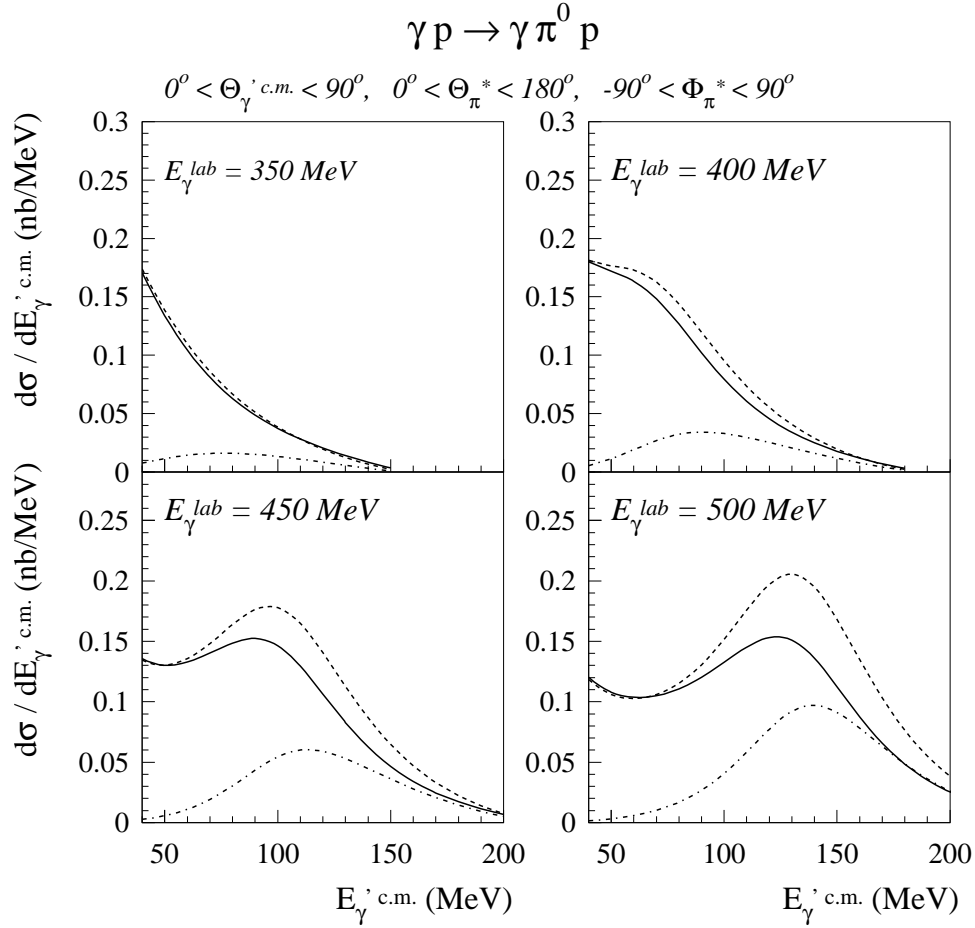}}
\caption[]{Outgoing photon c.m. energy dependence of the $\gamma \, p
  \, \to \, \gamma \, \pi^0 \, p$ cross section 
$d \sigma / d E_\gamma^{\, ' c.m.}$, partially integrated over the photon and 
pion angles as indicated. 
The calculations are the total results (including all
diagrams of Fig.~\ref{fig:gap_piogap_diag}) for the values~:
$\kappa_{\Delta +}$ = 0 (dashed curves) 
and $\kappa_{\Delta +}$ = 3 (full curves). For comparison, 
the calculation with only the $\Delta$ anomalous magnetic moment
contribution (a2) of Fig.~\ref{fig:gap_piogap_diag} for the value 
$\kappa_{\Delta +}$ = 3 is shown by the dashed-dotted curves. }
\label{fig:gap_ppioga_en3}
\end{figure}

\begin{figure}[ht]
\epsfxsize=13 cm
\centerline{\epsffile{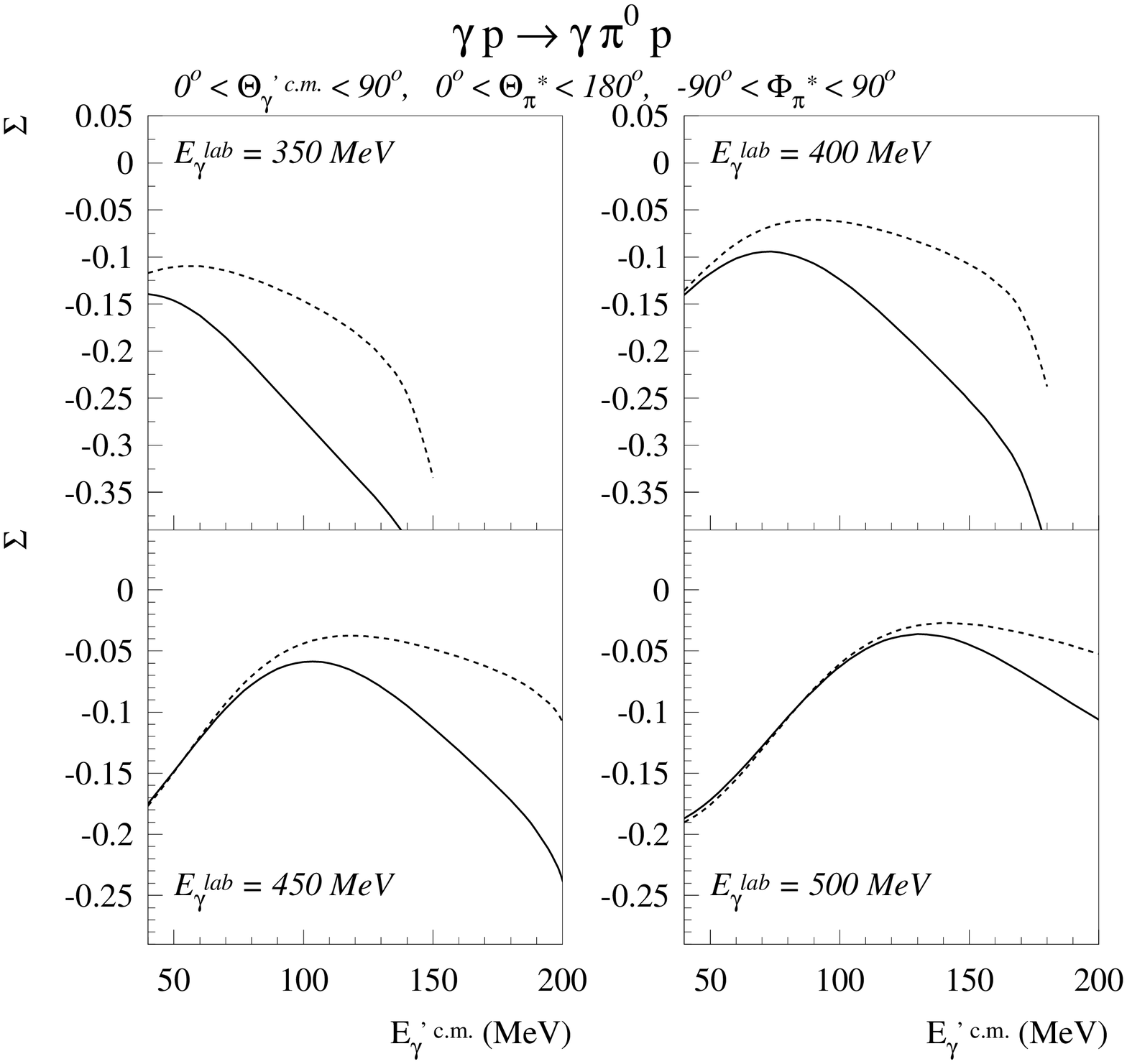}}
\caption[]{Outgoing photon c.m. energy dependence of the 
$\gamma \, p \, \to \, \gamma \, \pi^0 \, p$ 
photon asymmetry $\Sigma$, partially integrated over the photon and 
pion angles as indicated, and differential w.r.t. the outgoing photon energy 
$E_\gamma^{\, ' c.m.}$,
as in Fig.~\ref{fig:gap_ppioga_en3}. 
The calculations are the total results (including all
diagrams of Fig.~\ref{fig:gap_piogap_diag}), for the values~:
$\kappa_{\Delta +}$ = 0 (dashed curves) 
and $\kappa_{\Delta +}$ = 3 (full curves).}
\label{fig:gap_ppioga_en3_asymm}
\end{figure}

Having described the model for the $\gamma p \to \gamma \pi^0 p$ reaction, 
we now turn to the results. In order to have a point of comparison, we
start by showing the result for the $\gamma p \to \pi^0 p$ reaction
corresponding to the mechanisms of Fig.~\ref{fig:gap_piop_diag}. 
Our aim here is not to provide an accurate description of 
the $\gamma p \to \pi^0 p$ reaction (which exists in the literature), but
to see the intrinsic accuracy of the present effective Lagrangian
formalism, which serves as our starting point for the corresponding
mechanisms as shown in Fig.~\ref{fig:gap_piogap_diag} for the $\gamma p \to
\gamma \pi^0 p$ reaction. 
\newline
\indent
Fig.~\ref{fig:tot_ppio} shows the total cross section for the $\gamma
p \to \pi^0 p$ reaction. Although the $\gamma p \to \pi^0 p$ cross
section is clearly dominated
by the $\Delta$-excitation mechanism, making it an informative tool to
study the $\Delta$-resonance properties, additional non-resonant
mechanisms are needed to obtain an accurate description. 
It is seen that at the lower energies, 
the main non-resonant contributions come from 
the nucleon Born diagrams. At the higher energies, the
$\Delta$-excitation alone yields too large cross sections and is
mainly reduced by the $\omega$-exchange mechanism, which dominates the
high-energy behaviour of the $\gamma p \to \pi^0 p$ reaction \cite{GLV}.
By comparing the total result in Fig.~\ref{fig:tot_ppio} with the
accurate data, it is seen that the present effective Lagrangian
model provides a rather good description below the
$\Delta$-resonance position, underestimates the data around 325 MeV at
the 10~\% level, and has an accuracy at the 20~\% level around 450 MeV. 
The systematic deviation of the present tree-level description
with regard to the data as one moves through the $\Delta$-resonance, is well
understood and originates from $\pi N$ rescattering (loop) contributions
which restore unitarity. Although below two-pion production threshold, 
one can easily restore unitarity for the $\gamma p \to \pi^0 p$ reaction 
by using the $\pi N$ phase-shifts as phases of the pion
photoproduction amplitude,  
the corresponding unitarization procedure
for the $\gamma p \to \gamma \pi^0 p$ reaction has not yet been worked
out. In particular, for the $\gamma p \to \gamma \pi^0 p$ reaction, 
care has to be taken to respect gauge invariance, as one can have
intermediate $\pi^+ n$ loop contributions where the photon couples to
the intermediate charged pion. Such a gauge invariant unitary 
description of the $\gamma p \to \gamma \pi^0 p$ reaction 
could be addressed within the context of a dynamical model, 
and seems to be a promising subject for
future work. In the present paper, we explore as a first step the
$\gamma p \to \gamma \pi^0 p$ reaction within the context of the
effective Lagrangian model of Fig.~\ref{fig:gap_piogap_diag}, in order
to investigate promising kinematics and observables for the extraction
of the $\Delta^+$ magnetic dipole moment. 
\newline
\indent
Passing next from the $\gamma p \to \pi^0 p$ reaction to the
description of the $\gamma p \to \gamma \pi^0 p$ reaction, it is
important to stress that the electric coupling of the emitted 
photon to the charged particles 
is completely constrained by gauge invariance as discussed
in Sec.~\ref{model}. However the magnetic dipole moment of the
$\Delta^+$ (and also its smaller higher moments) is not fixed by
general theoretical arguments and has to be extracted from a fit to
experiment. Therefore, the only new parameter that appears in the
present formalism for the $\gamma p \to \gamma \pi^0 p$ reaction
compared to the $\gamma p \to \pi^0 p$ reaction, is the $\Delta^+$
anomalous magnetic moment $\kappa_{\Delta^+}$. In the following, we
investigate the sensitivity of different $\gamma p \to \gamma \pi^0 p$
observables to this anomalous magnetic moment $\kappa_{\Delta^+}$. 
\newline
\indent
Fig.~\ref{fig:gap_ppioga_en1} shows the dependence of the 
differential cross section of the $\gamma p \to \gamma \pi^0 p$ 
reaction on the outgoing photon energy
for three incoming photon energies through the $\Delta$-resonance region.   
The cross sections have been integrated over both photon and pion angles.
The calculations compare the importance of the different mechanisms in
Fig.~\ref{fig:gap_piogap_diag}, using the value $\kappa_{\Delta^+}$ = 3. 
One clearly sees that the cross section for the resonant mechanism
alone is dominated at the lower photon energies by the bremsstrahlung
contributions (Fig.~\ref{fig:gap_piogap_diag}(a1) and (a3)), which
show the typical 1/$E_\gamma^{'}$ energy dependence. It is also seen
that the non-resonant mechanisms induce, due to interference, 
a distinct energy dependence as one moves through 
the $\Delta$-resonance similar to the case of the $\gamma p \to \pi^0
p$ reaction. Note in particular the partial cancellation of
bremsstrahlung with $\Delta$ and nucleon intermediate states for the
smaller values of $E_\gamma^{'}$. 
\newline
\indent 
The cross section for the $\gamma p \to \gamma \pi^0 p$
reaction has been measured for the first time by the A2/TAPS
Collaboration at MAMI \cite{Kot99}. Our total results in 
Fig.~\ref{fig:gap_ppioga_en1} compare favorably with 
the preliminary data in both incoming and outgoing photon energy dependence.
\newline
\indent
Fig.~\ref{fig:gap_ppioga_en2} shows the sensitivity of the angle 
integrated cross sections to the value of the $\Delta^+$ anomalous
magnetic moment. The sensitivity of these cross sections to
$\kappa_{\Delta^+}$ increases with increasing outgoing photon energy,
but remains relatively modest even at the higher energies, where the
difference between the $\kappa_{\Delta^+} = 0$ and $\kappa_{\Delta^+} = 3$
values is at the 10 - 15 \% level. 
\newline
\indent
In Fig.~\ref{fig:gap_ppioga_ang_taps}, we show the corresponding
photon c.m. angular dependence of the cross sections integrated over
the outgoing photon energy (for the cut $E_\gamma^{\, ' c.m.} >
90$~MeV) and over the pion angles. 
The angular dependence displays a 
broad maximum around a c.m. angle of 110$^o$. 
The calculations show that the sensitivity of these integrated 
unpolarized cross sections to $\kappa_{\Delta^+}$ lies 
at the 10 - 20~\% level over most of the angular range. At an incoming photon
energy of 450 MeV, a larger effect is observed 
around the forward and backward angles. 
\newline
\indent
However, we found that the sensitivity to the $\Delta^+$ anomalous magnetic
moment can be substantially increased when using a linearly polarized
photon beam. In this case, one can define the photon asymmetry 
$\Sigma \equiv (d \sigma_\perp - d \sigma_\parallel)
/(d \sigma_\perp + d \sigma_\parallel)$, where $d \sigma_\perp$ ($d
\sigma_\parallel$) are the differential cross sections for incoming photon
polarization perpendicular (parallel) to the reaction plane spanned by
the incoming and outgoing photon c.m. momentum vectors. 
The outgoing photon energy dependence of the photon asymmetry 
is shown in Fig.~\ref{fig:gap_ppioga_en_asymm} 
where the differential cross
sections $d \sigma_\perp$ and $d \sigma_\parallel$ are integrated over
all photon and pion angles, corresponding to the situation in 
Fig.~\ref{fig:gap_ppioga_en2}. In comparison with the unpolarized cross
section shown in Fig.~\ref{fig:gap_ppioga_en2}, one now obtains a
clearly larger sensitivity to $\kappa_\Delta^+$, in particular at the
larger values of $E_\gamma^{\, ' c.m.}$.
\newline
\indent
One can further increase the sensitivity of the $\gamma p \to \gamma
\pi^0 p$ reaction to the $\Delta^+$ magnetic dipole moment by
measuring differential cross sections. 
In particular, by chosing those angular
ranges where the photon is emitted in the backward hemisphere 
with respect to both protons (in the c.m. system) as shown in 
Fig.~\ref{fig:kinem_forw}, one suppresses the bremsstrahlung
contributions and largely increases the sensitivity to $\kappa_{\Delta^+}$. 
This is shown in Fig.~\ref{fig:diffasymm4} 
for the fully differential cross section of Eq.~(\ref{eq:cross3b}), which is
differential w.r.t. the photon c.m. energy, the photon c.m. polar
angle and the pion polar and azimuthal angles (the latter two are
defined in the $\pi^0 p$ rest frame). The differential cross sections in 
Fig.~\ref{fig:diffasymm4}, shown for $E_\gamma$ = 450 MeV, 
are now peaked around 
$E_\gamma^{\, ' c.m.}$ = 100 MeV due to the resonant-$\Delta$
contribution of Fig.~\ref{fig:gap_piogap_diag}(a2), which contains
$\kappa_{\Delta^+}$. On the other hand,
in the angle integrated cross sections shown in 
Fig.~\ref{fig:gap_ppioga_en2}, the contribution of 
Fig.~\ref{fig:gap_piogap_diag}(a2) merely manifests itself in the shoulder
 at higher values of the outgoing photon energy $E_\gamma^{\, ' c.m.}$.
Furthermore, one sees from Fig.~\ref{fig:diffasymm4} that the maximum
of the cross section for $\kappa_{\Delta^+} = 0$ is reduced by about 30 \%
for the value $\kappa_{\Delta^+} = 3$ as compared to $\kappa_{\Delta^+} = 0$. 
The corresponding photon asymmetry $\Sigma$ reaches large values 
and also displays a very pronounced energy dependence. 
The differences resulting from various values of $\kappa_{\Delta^+}$
is in the range predicted by different theoretical models, 
and should be clearly resolved by measurements of $\Sigma$ at the 5 \% level. 
\newline
\indent
Although a measurement of the fivefold differential cross section is
very tough due to the small count rates, we found that one can largely
keep the sensitivity to $\kappa_\Delta$ by integrating over a larger
region of phase space around the kinematical situations where the
photon is emitted in the backward hemisphere with respect to both protons. 
This is illustrated in Fig.~\ref{fig:gap_ppioga_en3} where the
differential cross section is shown integrated over the forward
hemisphere for the outgoing photon angle (w.r.t. the incoming photon
direction in the c.m. system) and where the pion is emitted over 
the azimuthal angular range $-90^o \,<\, \Phi_\pi^* \,<\, 90^o$.
One sees from Fig.~\ref{fig:gap_ppioga_en3} that the cross section at
450 MeV shows a 20~\% sensitivity, and at 500 MeV a 30~\% sensitivity,
to the difference between $\kappa_{\Delta^+} = 0$ and $\kappa_{\Delta^+} = 3$. 
Furthermore, by comparing for a value $\kappa_{\Delta^+} = 3$, the full
result with the $\Delta$ anomalous magnetic moment contribution
separately (diagram Fig.~\ref{fig:gap_piogap_diag} (a2)), one sees that
the latter yields the dominant contribution to the total result at the
higher outgoing photon energies for 450 MeV and 500 MeV 
initial photon energies.
By comparing the partially integrated cross section of
Fig.~\ref{fig:gap_ppioga_en3} at 450 MeV to the cross section of 
Fig.~\ref{fig:gap_ppioga_en2}, one sees that it reaches about one
fifth of the cross section integrated over the full angular range. As
the cross section of Fig.~\ref{fig:gap_ppioga_en2} has already been
measured \cite{Kot99}, the cross sections of Fig.~\ref{fig:gap_ppioga_en3} 
could be accessible through a dedicated experiment.  
In view of the characteristic angular dependence of the effect under
investigation, a full $4 \pi$ coverage will be a prerogative to obtain
more significant information on the magnetic dipole moment of the
$\Delta$. Such a $4 \pi$ detector will allow one to increase the
count rate substantially w.r.t. the present TAPS
experiment, and thus to select kinematical ranges where background
(bremsstrahlung) effects are strongly suppressed as is shown in
Fig.~\ref{fig:gap_ppioga_en3}.  
\newline
\indent
Finally, we show in Fig.~\ref{fig:gap_ppioga_en3_asymm} the photon
asymmetries corresponding to the partially integrated cross sections
of Fig.~\ref{fig:gap_ppioga_en3}. Compared to the asymmetries of
Fig.~\ref{fig:gap_ppioga_en_asymm}, where the cross sections were
integrated over all photon and pion angles, both the magnitude of the
asymmetries in Fig.~\ref{fig:gap_ppioga_en3_asymm} and the sensitivity
to $\kappa_\Delta$ are larger. 
Therefore, a dedicated experiment consisting of a $4 \pi$ detector and
a high intensity polarized photon beam, promises to provide a good
opportunity to extract the $\Delta^+$ magnetic dipole moment
from the $\gamma p \to \gamma \pi^0 p$ reaction. Such an experiment is
planned using the Crystal Ball detector in combination with the polarized
photon beam at MAMI \cite{CB}.

\section{Conclusions}
\label{conclusions}
In this paper we investigated the $\gamma p \to \gamma \pi^0 p$
process as a tool to access the $\Delta^+(1232)$ magnetic dipole
moment. 
\newline
\indent
An effective Lagrangian formalism to describe the 
$\gamma p \to \pi^0 p$ reaction in the $\Delta(1232)$ resonance
region, served as a starting point to develop a corresponding 
model of the $\gamma p \to \gamma \pi^0 p$ reaction. 
When extending the model of the $\gamma p \to \pi^0 p$
process to the $\gamma p \to \gamma \pi^0 p$ process, the electric coupling of
the additional photon to the charged particles is completely
constrained by gauge invariance. In particular, we paid special
attention to the constraint that gauge invariance imposes on the
coupling of a photon to a resonance of finite width. In analogy to the
case of the $W^\pm$ vector boson, which has been discussed extensively
in recent years in the literature, we follow a complex mass scheme
procedure where the amplitude for the 
$\gamma p \to \gamma \pi^0 p$ process which involves the
$\Delta^+(1232)$ resonance, is Laurent expanded around the complex
pole position of the resonance. This procedure guarantees electromagnetic gauge
invariance. In comparison with the $\gamma p \to \pi^0 p$ reaction, 
the only new parameter which enters in the description of the $\gamma
p \to \gamma \pi^0 p$ reaction in the $\Delta(1232)$ region is the
$\Delta^+$ anomalous magnetic moment $\kappa_{\Delta^+}$.
We then investigated the sensitivity of various $\gamma p \to \gamma \pi^0 p$
observables to this anomalous magnetic moment $\kappa_{\Delta^+}$. 
\newline
\indent
We first compared the unpolarized $\gamma p \to \gamma \pi^0 p$ cross
section integrated over the whole photon and pion angular range. For
this fully angle integrated cross section, the resonant mechanism is
dominated by the bremsstrahlung contribution which shows the typical
$1/E_\gamma'$ dependence in the outgoing photon energy $E_\gamma'$. 
We also found that the non-resonant mechanisms induce, due to
interference, a distinct energy pattern as one moves through the
$\Delta$-resonance. Our calculations compare favorably with
preliminary data for those fully integrated cross sections. 
\newline
\indent
The sensitivity of the unpolarized $\gamma p \to \gamma \pi^0 p$ cross
section integrated over the whole photon and pion angular range to
$\kappa_{\Delta^+}$, was found to increase with increasing photon
energy, but remains relatively modest. At an incoming
photon energy of 450 MeV, the difference between the
$\kappa_{\Delta^+} = 0$ and $\kappa_{\Delta^+} = 3$ values is 
at the 10 - 15 \% level.  
\newline
\indent
However, one can further increase the sensitivity 
to $\kappa_{\Delta^+}$ by measuring differential cross sections. In
particular, by selecting the kinematical situations where the photon
is emitted in the backward hemisphere w.r.t. both protons, the
bremsstrahlung contribution to the cross sections is largely
suppressed. When partially integrating the cross sections around these
backward angle kinematics, we found that, for an incoming photon energy range
of 450 - 500 MeV, the difference  between the
$\kappa_{\Delta^+} = 0$ and $\kappa_{\Delta^+} = 3$ values is   
at the 20 - 30 \% level.  
\newline
\indent
Furthermore, we showed that the sensitivity 
to the $\Delta$ anomalous magnetic moment
can be even more increased by using a linearly polarized photon beam. At 
the higher outgoing photon energies, the asymmetries are at the 10 -
30 \% level through the $\Delta$ region and display a clear
sensitivity to $\kappa_{\Delta^+}$. 
\newline
\indent
In view of our findings, a dedicated experiment 
using a $4 \pi$ detector and a high intensity
photon beam seems to be a promising experiment to extract
the $\Delta^+$ magnetic dipole moment from the 
$\gamma p \to \gamma \pi^0 p$ reaction. 
The model developed in this
paper can serve as a first step for such an extraction. 
The measurement of the $\Delta$ magnetic dipole moment will provide 
us with new information on baryon structure 
which can be confronted with various baryon structure calculations.

\section*{Acknowledgments}

It is a pleasure to acknowledge fruitful conversations with
R. Beck, M. Kotulla, V. Metag, B. Nefkens, G. Rosner, and L. Tiator.
This work was supported by the Deutsche Forschungsgemeinschaft (SFB 443).

\newpage

\newpage

\newpage

\newpage

\newpage

\newpage

\newpage

\newpage

\newpage

\newpage

\begin{thebibliography}{00}

\bibitem{Ali00}
T.M. Aliev, A. \"Ozpineci, and M. Savci, 
Phys. Rev. D {\bf 62}, 053012 (2000).
\bibitem{Ali00b}
T.M. Aliev, A. \"Ozpineci, and M. Savci, 
Nucl. Phys. {\bf A678}, 443 (2000).


\bibitem{Kon68} 
L.A. Kondratyuk und L.A. Ponomarov, 
Yad. Fiz. {\bf 7}, 11 (1968) [Sov. J. Nucl. Phys. {\bf 7}, 82 (1968)].
\bibitem{Nef78}  
B.M.K. Nefkens {\it et al.}, Phys. Rev. D {\bf 18}, 3911 (1978). 
\bibitem{Bos91}  
A. Bosshard et al., Phys. Rev. D {\bf 44}, 1962 (1991).
\bibitem{PDG00}
D.E. Groom {\it et al.} (PDG), Eur. Phys. J. {\bf C15}, 1 (2000).
\bibitem{Dre83}
M.M. Giannini, in {\it Proceedings of the Workshop Perspectives on
  Nuclear Physics at Intermediate Energies, ICTP Trieste, Italy, 1983
  (World Scientific, Singapore, 1984)};  and 
D. Drechsel, in MAMI funding proposal to DFG, SFB 201 (1984-86), p. 56.
\bibitem{Mac99} 
A.I. Machavariani, A. Faessler, and A. J. Buchmann, 
Nucl. Phys. {\bf A646}, 231 (1999); Nucl. Phys. {\bf A686}, 601 (E) (2001).
\bibitem{DVGS} 
D. Drechsel, M. Vanderhaeghen, M.M. Giannini, and E. Santopinto,  
Phys. Lett. B {\bf 484}, 236 (2000).
\bibitem{Kot99}
M. Kotulla (for the A2/TAPS collaboration), in 
{\it Proceedings of the Workshop on The Physics of Excited 
Nucleons (Nstar 2001), Mainz, Germany, 2001, (World Schientific,
Singapore, to be published)}. 
\bibitem{CB}
R. Beck, B. Nefkens {\it et al.}, Letter of Intent, MAMI (2001).
\bibitem{Ami92}
M. El Amiri, G. L\'opez Castro, and J. Pestieau, 
Nucl. Phys. {\bf A543}, 673 (1992).
\bibitem{Cas00}
G. L\'opez Castro and A. Mariano, nucl-th/0010045.

\bibitem{Nath71}
L.M. Nath, B. Etemadi, and J.D. Kimel, Phys. Rev. D {\bf 3}, 2153 (1971).
\bibitem{MAID}
D. Drechsel, O. Hanstein, S. Kamalov, and L. Tiator, 
Nucl. Phys. {\bf A645}, 145 (1999).
\bibitem{Beck00}
R. Beck, H.P. Krahn {\it et al.}, Phys. Rev. C {\bf 61}, 035204 (2000).

\bibitem{Sir91}
A. Sirlin, Phys. Rev. Lett. {\bf 67}, 2127 (1991);
Phys. Rev. Lett. {\bf 86}, 389 (2001).
\bibitem{Stu93}
R.G. Stuart, Phys. Rev. Lett. {\bf 70}, 3193 (1993).
\bibitem{Vel94}
H. Veltman, Z. Phys. C {\bf 62}, 35 (1994). 
\bibitem{Lop95}
G. L\'opez Castro, J.L. Lucio M., and J. Pestieau, 
Int. J. Mod. Phys. A {\bf 11}, 563 (1996); hep-ph/9504351.
\bibitem{Baur95}
U. Baur and D. Zeppenfeld, Phys. Rev. Lett. {\bf 75}, 1002 (1995).
\bibitem{Beu97}
M. Beuthe, R. Gonzalez Felipe, G. L\'opez Castro, and J. Pestieau, 
Nucl. Phys. {\bf B498}, 55 (1997). 
\bibitem{Lop00}
G. L\'opez Castro and G. Toledo S\'anchez, 
Phys. Rev. D {\bf 61}, 033007 (2000).

\bibitem{GLV}
M. Guidal, J.M. Laget, and M. Vanderhaeghen, 
Nucl. Phys. {\bf A627}, 645 (1997).
\bibitem{vpi99}
M.M. Pavan, R.A. Arndt, I.I. Strakovsky, and R.L. Workman, 
$\pi$N Newslett. {\bf 15}, 171 (1999).

\bibitem{piotot_macc}
M. MacCormick et al., Phys. Rev. C {\bf 53}, 41 (1996). 
\bibitem{piotot_ahr}
J. Ahrens et al., Phys. Rev. Lett. {\bf 84}, 5950 (2000). 



\end{thebibliography}
\end{document}